\documentclass[12pt,leqno]{article}
\usepackage{amsmath,amsthm}

\def\init{\setcounter{equation}{0}}
\newtheorem{theorem}{Theorem}[section]
\newcommand{\R}{{\bf R}}

\newcommand{\Z}{{\bf Z}}

\newtheorem{lemma}[theorem]{Lemma}

\newtheorem{pro}{Proposition}[section]

\newcommand{\e}{{\varepsilon}}

\usepackage{tikz}

\title{Remarks on magnetic and electric  Aharonov-Bohm effects
\author{G.Eskin, \ \ \  Department of Mathematics, UCLA,\\ Los Angeles,
CA 90095-1555, USA. \ E-mail: eskin@math.ucla.edu}
}

\begin{document}

\maketitle

\begin{abstract}
We give a direct  proof of the magnetic  Aharonov-Bohm effects
without using the scattering theory and the theory of inverse boundary value
problems.  
This proof can serve as a framework  for a physical  experiment to confirm
the magnetic AB effect.  We prove also the electric AB effect and
we  suggest a physical experiment  to demonstrate the electric 
AB effect. 
In addition, 
 we   consider  a combined  electric and magnetic AB effect and
we propose a new inverse  problem for the time-dependent Schr\"odinger equations.
Finally we study the gravitational AB effect.
\end{abstract}

\section{Introduction.}
\label{section 1}
\init

Let $\Omega_1$  be a bounded domain in $\R^2$,  called the obstacle.  
  Consider the time-dependent 
 Schr\"odinger equation  in $(\R^2\setminus\overline \Omega_1)\times (0,T)$:
\begin{equation}                                       \label{eq:1.1}
-ih\frac{\partial u}{\partial t}+
\frac{1}{2m}\sum_{j=1}^n\left(-ih\frac{\partial}{\partial x_j}
-\frac{e}{c}A_j(x)\right)^2u+e V(x)u=0,
\end{equation}
where  $n=2$,
\begin{equation}                                           \label{eq:1.2}
u|_{\partial \Omega_1\times (0,T)}=0,
\end{equation}
\begin{equation}                                           \label{eq:1.3}
u(x,0)=u_0(x),\ \ x\in \R^2\setminus\Omega_1,
\end{equation}
and the electromagnetic potentials $A(x), V(x)$ are independent of $t$.
Let
\begin{equation}                                           \label{eq:1.4}
\alpha=\frac{e}{hc}\int_\gamma A(x)\cdot dx
\end{equation}
be the magnetic flux,  where $\gamma$  is a simple closed contour containing $\Omega_1$.
In seminal paper [AB] Y. Aharonov  and D. Bohm discovered  
that even if the magnetic field $B(x)=\mbox{curl\ }A=0$
in $\R^2\setminus \overline \Omega_1$,    
the magnetic potential $A$  has a physical impact in 
$\R^2\setminus\overline \Omega_1$  when  $\alpha\neq 2\pi n$,   $n$  is
an integer.  This phenomenon is called the Aharonov-Bohm  effect.  They proposed 
a physical experiment  to test this effect.  The experimental proof of AB effect  was
not easy to achieve.  The most ``clean"  AB experiment  was done by Tonomura 
et al [T et al].  A rigorous mathematical justification of the Tonomura et al 
experiment   was given  in [BW2],  [BW3].

In the same paper [AB] Aharonov and Bohm gave a mathematical proof of AB effect by 
showing that 
the scattering cross section 
 depends  on $\alpha$.
The proofs of AB effects  using the scattering theory and the inverse scattering were given 
also in  [R], [N], [W1], [BW1], [RY1], [RY2], [Y], [EI], [EI0].  
The inverse boundary value problems
approach to AB effect  was developed in [E1], [E2], [E3], [E4], [E5].
Note that
the solution of the inverse scattering problem can be reduced to the solution 
of the inverse boundary value problem.  This reduction is well-known in the case 
$n\geq 3$  and  electromagnetic fields with compact supports (see,  for example, [E5], \S1).
In the case $n=2$ the reduction was proven in [EIO].  
Note that  the class of electromagnetic fields with  compact  supports is the
natural setting  for the study of AB effect.

In this paper we give a new direct mathematical proof of the magnetic AB  effect
that uses the relation between the solutions of the Schr\"odinger  equation and
the wave equation (see [K]).   This proof  can be used
as a framework  for a physical experiment to verify  the magnetic AB effect.
We consider  the case of one and several obstacles.  The case of several obstacles requires a technique of broken  rays.  It allows to detect  the magnetic AB effect
in the case when obstacles are close to each other and the treatment
of the cluster of obstacles as one obstacle may miss the AB effect.
We give also a rigorous proof of the electric AB effect and propose a physical 
experiment to verify it.
We show that the electric AB effect occurs only when the domain 
is time-dependent  with its topology changing in time.  In addition 
 we consider  combined electric and magnetic AB effect,  and gravitational AB effect 

The plan of the paper is the following:

In \S 2 we state the AB effect.  In \S 3  we prove the magnetic AB  
effect
in the case of one obstacle in two and three dimensions  and in the case of
several obstacles.
 In \S 4 we
consider the electric AB effect,  and in \S 5 the combined electric and magnetic AB effect
for time-dependent electromagnetic potentials.
In the end of \S 5  we study a new inverse problem for the time-dependent
Schr\"odinger equations.  In \S 6  we prove a general case of the gravitational AB effect.
A particular case was considered previously in [S].

\section{The magnetic AB effect}
\label{section 2}
\init

Let $\Omega_1,...,\Omega_m$  be smooth obstacles in $\R^n$.  Assume 
that
$\overline \Omega_j\cap\overline \Omega_k\neq 0$ if $j\neq k$.
Consider the  Schr\"odinger equation (\ref{eq:1.1})  in 
$(\R^n\setminus\overline\Omega)\times (0,T), \ n\geq 2$,  where 
$\Omega=\cup_{j=1}^m\Omega_j$,
\begin{equation}                                 \label{eq:2.1}
u\big|_{\partial\Omega\times(0,T)}=0,
\end{equation} 
and (\ref{eq:1.3}) holds in $\R^n\setminus\Omega$.

In (\ref{eq:1.1})  $A(x)=(A_1(x),...,A_n(x))$ is the magnetic 
potential and $V(x)$  is the electric potential.

In this paper we assume that $B(x)=\mbox{curl}\ A(x)=0$  in $\R^n\setminus\overline\Omega$,  
i.e.  the  magnetic field $B(x)$ is shielded inside $\Omega$.  For  the simplicity 
we assume  that $V(x)$  has a compact support.

Denote by  $G(\R^n\setminus \Omega)$  the group of $C^\infty$ complex-valued
functions $g(x)$ such that $|g(x)|=1$ in $\R^n\setminus\Omega$ 
and
\begin{eqnarray}
\nonumber
g(x)=1+O\Big(\frac{1}{|x|}\Big)\ \ \mbox{for}\ \ |x|>R\ \ \mbox{if}\ \ n\geq 3,
\\
\nonumber
g(x)=e^{ip\theta(x)}\bigg(1+O\Big(\frac{1}{|x|}\Big)\bigg)\ \ \mbox{for}\ \ |x|>R\ \ 
\mbox{if}\ \ n=2.
\end{eqnarray}
Here $p$ is an arbitrary integer, $0\in\Omega$  and $\theta(x)$  is
the polar angle  of $x$.  We call
$G(\R^n\setminus \Omega)$  the gauge group.   If  $u'(x)=g^{-1}(x)u(x)$  then
$u'(x)$  satisfies the Schr\"odinger
equation (\ref{eq:1.1})   with electromagnetic potentials $(A'(x),V'(x))$  where
\begin{eqnarray}                                       \label{eq:2.2}
V'(x)=V(x),
\\
\nonumber
\frac{e}{c}A'(x)=\frac{e}{c}A(x)+ihg^{-1}(x)\frac{\partial g(x)}{\partial x}.
\end{eqnarray}
We shall call electromagnetic potentials $(A',V')$  and  $(A,V)$  gauge equivalent   if there exists $g(x)\in G(\R^n\setminus\Omega)$
such that (\ref{eq:2.2})  holds.

We shall describe all gauge equivalence classes  
of potentials when  $B=\mbox{curl}\ A=0$  in  $\R^n\setminus\Omega$.

Consider   first  the case of the obstacle $\Omega_1$  in $\R^2$.  The gauge group
$G(\R^2\setminus\Omega_1)$  consists of $g(x)=e^{ip\theta(x)+\frac{i}{h}\varphi(x)}$,
where $p$  is an integer,  and
$\varphi(x)\in C^\infty(\R^2\setminus\Omega_1),\ 
\varphi(x)=O(\frac{1}{|x|})$  when $|x|>R$.
The gauge equivalence class is determined  by the magnetic flux (\ref{eq:1.4}) 
modulo $2\pi p,\ p\in\Z$.

In the case of several obstacles $\Omega_1,...,\Omega_m$  in $\R^2$  denote
by $\gamma_j,\ 1\leq j\leq m,$  a simple closed curve encircling $\Omega_j$  only.
Let
$$
\alpha_j=\frac{e}{hc}\int_{\gamma_j}A\cdot dx
$$
be the corresponding  magnetic flux.  Then numbers 
$\alpha_j(\mbox{mod}\ 2\pi n), j=1,...,m,$  
determine the gauge equivalent class of $(A,V)$.  

Finally,  in the case  of $m\geq 1$  in $\R^n,\ n\geq 3$,  it can be shown that there exists a finite number of closed curves $\gamma_1,...,\gamma_r$ in 
$\R^n\setminus\overline\Omega$ ($r=0$ if
$\R^n\setminus\Omega$  is simply-connected)  such that $(A,V)$  and  $(A',V')$  are
gauge equivalent iff 
$\frac{e}{hc}\int_{\gamma_j}A\cdot dx-
\frac{e}{hc}\int_{\gamma_j}A'\cdot dx=2\pi n_j,\ n_j\in\Z$,  
for all  $1\leq j\leq r$.
\qed

Any  two  electromagnetic potentials belonging  to the same gauge equivalence class represent the same physical reality  and can not be distinguished in any physical experiment.

The Aharonov-Bohm effect is the statement that electromagnetic potentials belonging to different gauge equivalence classes have a different physical impact.  Consider, 
for example,  the probability density $|u(x)|^2$.
It has the same value for any representative of the same gauge equivalence class since 
$|g^{-1}(x)u(x)|^2=
|u(x)|^2$.

To prove the AB effect it is enough to show that $|u(x)|^2$  changes 
for some $u(x)$ when we change 
the gauge equivalence class.

\section{The proof of the magnetic AB effect}
\label{section 3}
\init

\subsection{The case of one obstacle in $\R^2$}
Consider the Schr\"odinger equation (\ref{eq:1.1})   
in $(\R^2\setminus \Omega_1)\times(0,T)$  with the boundary condition (\ref{eq:1.2}) and the
initial condition (\ref{eq:1.3}).

Let 
$w(x,t)$  be the solution of the wave equation
\begin{equation}                                    \label{eq:3.1}
\frac{h^2}{2m}\frac{\partial^2 w}{\partial t^2}+H w=0\ \ \mbox{in}\ \ 
(\R^2\setminus \Omega_1)\times(0,+\infty)
\end{equation}
with the boundary condition
\begin{equation}                                       \label{eq:3.2}
w\big|_{\partial\Omega_1\times(0,+\infty)}=0
\end{equation}
and the initial conditions 
\begin{equation}                                      \label{eq:3.3}
w(x,0)=u_0(x),\ \  \frac{\partial w(x,0)}{\partial t}=0,\ \ x\in \R^2\setminus\Omega_1,
\end{equation}
i.e. $w(x,t)$ is even in $t$.
Here 
$$
H=\frac{1}{2m}\Big(-ih\frac{\partial}{\partial x}-\frac{e}{c}A\Big)^2+eV(x).
$$
There is a formula relating $u(x,t)$  and $w(x,t)$  (cf. [K]):
\begin{equation}                                        \label{eq:3.4}
u(x,t)=\frac{e^{-i\frac{\pi}{4}}\sqrt m}{\sqrt{2\pi ht}}
\int_{-\infty}^\infty e^{\frac{imx_0^2}{2ht}}w(x,x_0)dx_0.
\end{equation}
We shall consider  solutions of (\ref{eq:3.1}) such that
\begin{equation}                                       \label{eq:3.5}
|w(x,t)|\leq C(1+|t|)^m,\ \ 
\Big|\frac{\partial^r w(x,t)}{\partial t^r}\Big|
\leq C_r(1+|t|)^m,\ \ \forall r\geq 1.
\end{equation}
Let $\chi_0(t)\in C_0^\infty(\R^1),\ \chi_0(-t)=\chi_0(t),\ \chi_0(t)=1$
for  $|t|<\frac{1}{2},\ \chi_0(t)=0$   for  $|t|>1$  We define  the integral
(\ref{eq:3.4})   as the limit of
\begin{equation}                                        \label{eq:3.6}
\frac{e^{-i\frac{\pi}{4}}\sqrt m}{\sqrt{2\pi ht}}
\int_{-\infty}^\infty \chi_0(\e x_0) e^{\frac{imx_0^2}{2ht}}w(x,x_0)dx_0
\end{equation}
as $\e\rightarrow 0$,  and we shall show that this limit exists  for any
$w(x,x_0)$ satisfying  (\ref{eq:3.5}).  Substitute the identity
$$
\Big(\frac{ht}{imx_0}\frac{\partial}{\partial x_0}\Big)^M
 e^{\frac{imx_0^2}{2ht}}=e^{\frac{imx_0^2}{2ht}},
\ \ \ \forall M,
$$
in (\ref{eq:3.6}) and integrate by parts in (\ref{eq:3.6}) for  $|x_0|>1$.
If $M\geq m+2$  we get an absolutely integrable function of $x_0$ and 
therefore we can pass to the limit when $\e\rightarrow 0$.

Note that
\begin{equation}                                    
\nonumber
\Big(-ih\frac{\partial}{\partial t}+H\Big)u(x,t)
=\frac{e^{-i\frac{\pi}{4}}\sqrt m}{\sqrt{2\pi ht}}
\int_{\infty}^\infty e^{\frac{imx_0^2}{2ht}}
\Big(\frac{h^2}{2m}\frac{\partial^2}{\partial x_0^2}+H\Big)
w(x,x_0)dx_0,
\end{equation}
 Note also that
\begin{equation}
\nonumber
u(x,0)=\lim_{t\rightarrow 0}\frac{e^{-i\frac{\pi}{4}}\sqrt m}{\sqrt{2\pi ht}}
\int_{-\infty}^\infty e^{\frac{imx_0^2}{2ht}}w(x,x_0)dx_0=
w(x,0).
\end{equation}
Therefore $u(x,t)$  satisfies (\ref{eq:1.1}),  (\ref{eq:1.2}),  (\ref{eq:1.3})
if  $w(x,t)$  satisfies  (\ref{eq:3.1}),  (\ref{eq:3.2}),  (\ref{eq:3.3}).
\qed

We  shall  construct  geometric optics type solutions of
(\ref{eq:3.1}), (\ref{eq:3.2}), (\ref{eq:3.3}) and then use the formula 
(\ref{eq:3.4})  to obtain solutions of the Schr\"odinger equation.

We shall look for $w(x,t)$  in the form   
\begin{equation}                                \label{eq:3.7}
w_N(x,t)=e^{i\frac{mk}{h}(x\cdot\omega - t)}
\sum_{p=0}^N\frac{a_p(x,t)}{(ik)^p}+e^{i\frac{mk}{h}(x\cdot \omega+t)}
\sum_{p=0}^N\frac{b_p(x,t)}{(ik)^p},
\end{equation}
where $k$  is a large parameter.

Substituting
(\ref{eq:3.7})  into (\ref{eq:3.1})  and equating equal  powers of $k$  we get
\begin{eqnarray}                                     \label{eq:3.8}
ha_{0t}(x,t)+h\omega\cdot a_{0x}(x,t)-i\omega\cdot \frac{e}{c}A(x)a_0=0,
\\
\nonumber
-hb_{0t}(x,t)+h\omega\cdot b_{0x}-i\omega\cdot \frac{e}{c} A(x)b_0=0,
\end{eqnarray}
\begin{eqnarray}                                     \label{eq:3.9}
\ \ 
ha_{pt}(x,t)+h\omega\cdot a_{px}(x,t)-i\omega\cdot \frac{e}{c}A(x)a_p=
i\Big(\frac{h^2}{2m}\frac{\partial^2}{\partial t^2}+H\Big)a_{p-1},
\\
\nonumber
-hb_{pt}(x,t)+h\omega\cdot b_{px}-i\omega\cdot \frac{e}{c} A(x)b_p=
i\Big(\frac{\partial^2}{\partial t^2}+H\Big)b_{p-1},\ \ 1\leq p\leq N.
\end{eqnarray}
We have $b_p(x,t)=a_p(x,-t)$  for $p\geq 0$,  assuming that $b_p(x,0)=a_p(x,0)$.

Introduce new coordinates $(s,\tau,t)$ instead  of $(x_1,x_2,t)$  where 
\begin{eqnarray}                                 \label{eq:3.10}
s=(x-x^{(0)})\cdot\omega -t,
\\
\nonumber
\tau=(x-x^{(0)})\cdot \omega_\perp,
\\
\nonumber
t=t.\ \ \ \ \ \ \ \ \ \ \ \ \ \ \ \ \ \  
\end{eqnarray}
Here
$\omega_\perp\cdot \omega=0,\ |\omega_\perp|=|\omega|=1$.  We assume that  
$x^{(0)}$  is a fixed point outside of the obstacle $\Omega_1$  
and that the line $x=x^{(0)}+s\omega,\ s\in \R,$  does not intersect $\Omega_1$.
Equations (\ref{eq:3.8}), (\ref{eq:3.9})  have the following form in the new
coordinates
\begin{eqnarray}                                         \label{eq:3.11}
\hat a_{0t}(s,\tau,t)
-i\omega\cdot\frac{e}{hc}A(x^{(0)}+(s+t)\omega+\tau\omega_\perp)\hat a_0=0,
\\
\nonumber
\hat a_{pt}(s,\tau,t)
-i\omega\cdot\frac{e}{hc}A(x^{(0)}+(s+t)\omega+\tau\omega_\perp)\hat a_p=
\hat f_p(s,\tau,t),\ \ p\geq 1,
\end{eqnarray}
where 
$\hat a_p(s,\tau,t)=a_p(x,t),\ \hat f_p(s,\tau,t)$ is 
$\frac{i}{h}\Big(\frac{h^2}{2m}\frac{\partial^2}{\partial t^2}+H\Big)a_{p-1}$
in the 
new coordinates.

We impose the following initial conditions
\begin{eqnarray}                                        \label{eq:3.12}
\hat a_0(s,\tau,0)=\frac{1}{2}\chi_0\Big(\frac{\tau}{\delta_1}\Big)
\chi_0\Big(\frac{s}{\delta_2 k}\Big),
\\
\nonumber
\hat a_p(s,\tau,0)=0\ \ \mbox{for}\ \ p\geq 1,
\end{eqnarray}
where 
$\chi_0(s)$  is the same as above.  We assume  that $\delta_1$  is  such that  
$\mbox{supp\ }\chi_0(\frac{(x-x^{(0)})\cdot\omega_\perp}{\delta_1})$  does not 
intersect $\Omega_1$.
Then 
$$
\hat a_0(s,\tau,t)=\frac{1}{2}\chi_0\Big(\frac{\tau}{\delta_1}\Big)\chi_0
\Big(\frac{s}{\delta_2k}\Big)
\exp\Big(\frac{ie}{hc}\int_0^t\omega\cdot A(x^{(0)}+(s+t')
\omega+\tau\omega_\perp)dt'\Big).
$$
Since $s=(x-x^{(0)})\cdot \omega-t$  we have  in the original coordinates 
\begin{multline}                                         \label{eq:3.13}
a_0(x,t)=\frac{1}{2}\chi_0\Big(\frac{(x-x^{(0)})\cdot \omega_\perp}{\delta_1}\Big)
\chi_0\Big(\frac{(x-x^{(0)})\cdot \omega-t}{\delta_2k}\Big)
\\
\cdot
\exp\Big(\frac{i e}{hc}\int_0^t\omega\cdot A(x-t''w)dt''\Big),
\end{multline}
where we made the change of variables $t-t'=t''$.
Note  that
\begin{equation}                                    \label{eq:3.14}
|a_p(x,t)|\leq C t^p,\ \ 1\leq p\leq N,
\end{equation}
and (\ref{eq:3.5})  holds for any  $r\geq 1$.
Since  $b_p(x,t)=a_p(x,-t),\ p\geq 0,$   we
have that
\begin{eqnarray}                                         \label{eq:3.15}
\ \ \ \ \ \ \ 
w_N(x,0)=\chi_0\Big(\frac{(x-x^{(0)})\cdot \omega_\perp}{\delta_1}\Big)
\chi_0\Big(\frac{(x-x^{(0)})\cdot \omega}{\delta_2k}\Big)
e^{i\frac{m}{h}k\omega\cdot  x},
\\
\nonumber
w_{Nt}(x,0)=0.
\end{eqnarray}
Let 
\begin{equation}                                      \label{eq:3.16}
u_N(x,t)=\frac
{e^{-i\frac{\pi}{4}}\sqrt{m}}{\sqrt{2\pi ht}}\int_{-\infty}^\infty
e^{\frac{imx_0^2}{2ht}}w_N(x,x_0)dx_0.
\end{equation}
Using that $b_p(x,t)=a_p(x,-t)$ and making a change of variables we get
\begin{equation}                                       \label{eq:3.17}
u_N(x,t)=
\frac
{2e^{-i\frac{\pi}{4}}\sqrt{m}}{\sqrt{2\pi ht}}\int_{-\infty}^\infty
e^{\frac{imx_0^2}{2ht}+\frac{imk}{h}(x\cdot \omega-x_0)}
\sum_{p=0}^N\frac{a_p(x,x_0)}{(ik)^p}
dx_0.
\end{equation}
We have
\begin{equation}                                      \label{eq:3.18}
\Big(-ih\frac{\partial }{\partial t} +H\Big)u_N(x,t)=\frac
{e^{-i\frac{\pi}{4}}\sqrt{m}}{\sqrt{2\pi ht}}\int_{-\infty}^\infty
e^{\frac{imx_0^2}{2ht}}
\Big(\frac{h^2}{2m}\frac{\partial^2}{\partial x_0^2}+H\Big)
w_N(x,x_0)dx_0.
\end{equation}
Note that
\begin{multline}         
\Big(\frac{h^2}{2m}\frac{\partial^2}{\partial x_0^2}+H\Big)
w_N(x,x_0)=
e^{\frac{imk}{h}(x\cdot \omega-x_0)}
\Big(\frac{h^2}{2m}\frac{\partial^2}{\partial x_0^2}+H\Big)
a_N(x,x_0)
\\
\nonumber
+
e^{\frac{imk}{h}(x\cdot \omega+x_0)}
\Big(\frac{h^2}{2m}\frac{\partial^2}{\partial x_0^2}+H\Big)
b_N(x,x_0).
\end{multline}
Denote by $g_N(x,t)$ the  right hand  side  of (\ref{eq:3.18}).
Since $b_N(x,x_0)=a_N(x,-x_0)$  we have 
\begin{equation}                               \label{eq:3.19}
g_N(x,t)=
\frac 
{2e^{-i\frac{\pi}{4}}\sqrt{m}}{\sqrt{2\pi ht}}\int_{-\infty}^\infty
e^{\frac{imx_0^2}{2ht}}e^{\frac{imk}{h}(x\cdot\omega - x_0)}
\Big(\frac{h^2}{2m}\frac{\partial^2}{\partial x_0^2}+H\Big)
\frac{a_N(x,x_0)}{(ik)^N}dx_0.
\end{equation}
We apply 
the stationary phase method to the integrals (\ref{eq:3.17}).
The equation for the critical point is 
$\frac{mx_0}{ht}-\frac{mk}{h}=0$,  i.e.  $x_0=kt$  and the Hessian is  
$\frac{m}{ht}$.  Therefore
\begin{multline}                                \label{eq;3.20}
u_N(x,t)=e^{-\frac{imk^2t}{2h}+i\frac{mk}{h}x\cdot\omega}
\chi_0\Big(\frac{(x-x^{(0)})\cdot\omega_\perp)}{\delta_1}\Big)
\exp\Big(\frac{ie}{hc}\int_0^\infty\omega\cdot A(x-s'\omega)ds'\Big)
\\
+O(\e),
\end{multline}
where
$\e$  is arbitrary small when $k$  is sufficiently large and
$t$  is sufficiently small.

We used that $\chi_0\big(\frac{x\cdot \omega -k t}{\delta_2 k}\big)=1$
when $k$  is large and $t$  is small.
Note that 
\begin{equation}                              \label{eq:3.21}
\Big|\frac{a_p(x,kt)}{(ik)^p}\Big|
\leq \frac{C}{k^p}(kt)^p\leq Ct^p
\end{equation}is small when $t$  is small.

Applying the stationary phase method to (\ref{eq:3.19}) we get,  using (\ref{eq:3.14})
that 
\begin{equation}                                \label{eq:3.22}
\int_{\R^2\setminus\Omega}|g_N(x,t)|^2dx\leq C t^Nk^{\frac{1}{2}}.
\end{equation}We used in (\ref{eq:3.22})  that 
$\chi_0\big(\frac{x\cdot \omega -k t}{\delta_2 k}\big)=0$
when  $|x\cdot \omega|>Ck$.

Denote  by  $\|g_N\|_r$  the Sobolev norm in $H_r(\R^2\setminus\Omega_1)$.
It follows from (\ref{eq:3.19})  and (\ref{eq:3.14})  that 
$$
\|g_N\|_r\leq Ct^Nk^{r+\frac{1}{2}}.
$$
Let $R_N(x,t)$  be the solution of 
$$
\Big(-ih\frac{\partial}{\partial t}+H\Big)R_N=-g_N(x,t)
\ \ \mbox{in} \ \ (\R^2\setminus\Omega_1)\times(0,T),
$$
$$
R_N\Big|_{\partial\Omega_1\times(0,T)}=0,
$$
$$
R_N(x,0)=0.
$$
Such solution exists and satisfies the following estimates (cf. [E1]):
\begin{equation}                                   \label{eq:3.23}
\max_{0\leq t\leq T}\|R_N(\cdot,t)\|_3
\leq C\int_0^T\Big( \|g_N(\cdot,t)\|_1
+\big\|\frac{\partial g_N(\cdot,t)}{\partial t}\big\|_0\Big)dt.
\end{equation}
By the Sobolev embedding theorem
 $|R_N(x,t)|\leq C\max_{0\leq t\leq T}\|R_N(\cdot,t)\|_3$ for all
$(x,t)\in (\R^2\setminus\Omega_1)\times(0,T)$.  Since
$$
\max_{0\leq t\leq T} \Big\|\frac{\partial^p}{\partial t^p}g_N(x,t)\Big\|_r\leq 
CT^Nk^{\frac{1}{2}+r}
$$
we get that 
$$
|R_N(x,t)|\leq C\e
$$
if $T\leq \frac{C}{k^{\delta_3}},\  0<\delta_3<1, \ N\delta_3>\frac{3}{2}.
$

Note that 
$u=u_N+R_N$  satisfies (\ref{eq:1.1}), (\ref{eq:1.2}) and the initial condition
$u(x,0)=u_N(x,0)=e^{ik\omega\cdot x}\chi_0(\frac{(x-x^{(0)})\cdot\omega_\perp}{\delta_1})$
when $k$ is large.  Therefore we constructed a solution $u(x,t)$  for
$x\in \R^2\setminus\Omega_1,\ t\in (0,T),\ T=O(\frac{1}{k^\delta_3}),\ k$
is large,  such that
\begin{multline}                                       \label{eq:3.24}
u(x,t)
\\
=e^{-i\frac{mk^2t}{2h}-i\frac{mk}{h}x\cdot\omega}
\chi_0\Big(\frac{(x-x^{(0)})\cdot\omega_\perp}{\delta_1}\Big)
\exp\Big(i\frac{e}{hc}\int_0^\infty \omega\cdot A(x-s'\omega)ds'\Big)
\\
+O(\e),
\end{multline}
where $\e$  can be chosen arbitrary small  if  $k$  is large enough.
\qed

Let $x^{(0)}\in \R^2\setminus\Omega_1$  and let $\omega$  and $\theta$  be
two unit vectors (see Fig.1):

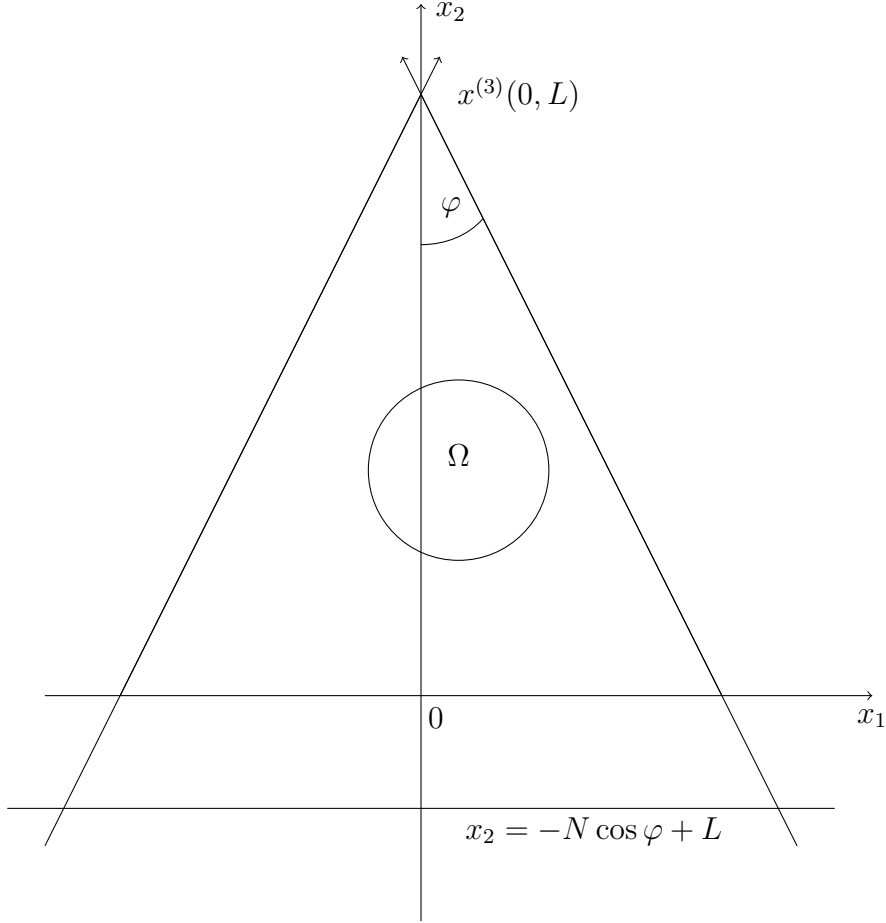
\begin{tikzpicture}

\draw[->](-5,0)--(6,0);
\draw[->](0,-3)--(0,9.2);

\draw[->](-5,-2)--(0.25,8.5);
\draw[->](5,-2)--(-0.25,8.5);

\draw(-5.5,-1.5)--(5.5,-1.5);

\draw(0,8)--(4,0);
\draw(0,8)--(-4,0);

\draw (0.5,3) circle (1.2 cm);
\draw (0.5,3.2) node {$\Omega$};
\draw(1.3,8) node {$x^{(3)}(0,L)$};

\draw(0.2,-0.3)  node {$0$};
\draw(2.3,-1.8)  node {$x_2=-N\cos\varphi +L$};
\draw(6,-0.3) node  {$x_1$};
\draw(0.4,9.1) node {$x_2$};

\draw(0,6) arc (270:325:1 and 0.8);
\draw (0.4,6.5) node {$\varphi$};

\end{tikzpicture}

 Figure 1.
\\
\\

Consider the difference 
of two solutions of the form (\ref{eq:3.24}) corresponding to
$(x^{(0)},\omega)$  and  $(x^{(0)},\theta)$,  respectively:
\begin{equation}                               \label{eq:3.25}
u=v_1(x,t,\omega)-v_2(x,t,\theta),
\end{equation}
\begin{equation}                                       \label{eq:3.26}
v_1(x,t,\omega)=e^{-i\frac{mk^2t}{2h}+i\frac{mk}{h}x\cdot\omega}
\chi_0\Big(\frac{(x-x^{(0)})\cdot\omega_\perp}{\delta_1}\Big)
\exp\Big(i\frac{e}{hc}\int_0^\infty \omega\cdot A(x-s'\omega)ds'\Big)+O(\e),
\end{equation}
\begin{equation}                                       \label{eq:3.27}
v_2(x,t,\theta)=e^{-i\frac{mk^2t}{2h}+\frac{mk}{h}x\cdot\theta}
\chi_0\Big(\frac{(x-x^{(0)})\cdot\theta_\perp}{\delta_1}\Big)
\exp\Big(i\frac{e}{hc}\int_0^\infty \theta\cdot A(x-s'\theta)ds'\Big)+O(\e),
\end{equation}
where $\theta_\perp\cdot\theta=0$.  Note that
modulo $O(\e)$  the support of $v_1$  is contained  in a small neighborhood 
of the line $x=x^{(0)}+s\omega$  and the support of $v_2$  is contained in
a small neighborhood of $x=x^{(0)}+s\theta$.

Let $U_0$  be a disk of radius $\e_0$  contained in
$(\mbox{supp\ }v_1)\cap(\mbox{supp\ }v_2)$.
We assume that $\chi_0(\frac{(x-x^{(0)})\cdot\omega_\perp}{\delta_1})=
\chi_0(\frac{(x-x^{(0)})\cdot\theta_\perp}{\delta_1})=1$  
in $U_0$. We have for $x\in U_0$
and $0<t<T=\frac{1}{k^{\delta_3}}$
\begin{multline}                                              \label{eq:3.28}
|v_1(x,t)-v_2(x,t)|^2=
\Big|1-e^{i\frac{mk}{h}x\cdot(\omega-\theta)+i(I_1-I_2)}\Big|^2+O(\e)
\\
=4\sin^2\frac{1}{2}\Big(\frac{mk}{h}x\cdot(\omega-\theta)+I_1-I_2\Big)+O(\e),
\\
I_1=\frac{e}{hc}\int_0^\infty\omega\cdot A(x-s\omega)ds,\ \ \
I_2=\frac{e}{hc}\int_0^\infty\theta\cdot A(x-s\theta)ds,
\end{multline}    
and $k>k_0,\ k_0$  is large,  $T\leq \frac{1}{k_0^{\delta_3}}$.

Choose $k_n>k_0$  such  that
\begin{equation}                                       \label{eq:3.29}
\frac{mk_n}{h}x^{(0)}\cdot(\omega-\theta)=2\pi n,\ \ n\in\Z.
\end{equation}
Let, for simplicity,  $\theta_1=\omega_1,\ \theta_2=-\omega_2,\ x^{(0)}=(0,L),\ 
\tan\varphi=\frac{\theta_2}{\theta_1}$  is small.  Define 
$$
I_{1N}(x,\omega)=
\frac{e}{hc}\int_0^{N}\omega\cdot A(x^{(0)}-s\omega)ds,
$$
$$
I_{2N}(x,\theta)=\frac{e}{hc}\int_0^N\theta\cdot A(x^{(0)}-s\theta)ds,
$$
$$
I_{3N}=\frac{e}{hc}\int_{-N\sin\varphi}^{N\sin\varphi}A_1(s,-N\cos\varphi+L)ds
$$
(see Fig.1).
Note 
that $I_{1N}(x^{(0)},\omega)-I_{2N}(x^{(0)},\theta)+I_{3N}=\alpha$,
where $\alpha$ is the magnetic flux  (cf. (\ref{eq:1.4})).   We assume that
\begin{equation}                                        \label{eq:3.30}
\alpha\neq 2\pi n,\ \forall n\in \Z.
\end{equation} 
Since $|A|\leq\frac{C}{r}$,  where  $r$
is the distance to $\Omega_1$,
we have
\begin{equation}                                       \label{eq:3.31}
|I_{3N}|\leq\frac{e}{hc}\frac{C}{N}2N\sin\varphi=
C_1\frac{e}{hc}\sin\varphi.
\end{equation}
When $N\rightarrow\infty$  we get
\begin{equation}                                        \label{eq:3.32}
I_1-I_2=\alpha+O\Big(\frac{e}{hc}\sin\varphi\Big)\ \ \mbox{for}\ \ x\in U_0.
\end{equation}
Assuming that the radius of the disk $U_0$  is $\e_0$  we get
\begin{equation}                                      \label{eq:3.33}
\Big|\frac{mk_n}{h}(x-x^{(0)})\cdot(\omega-\theta)\Big|\leq C\frac{mk_n}{h}\e_0\sin\varphi.
\end{equation}
Therefore using (\ref{eq:3.29}),  (\ref{eq:3.31}),  (\ref{eq:3.32}),  (\ref{eq:3.33}),
fixing  $k_n>k_0$  and  choosing  $\varphi$  and  $\e_0$  small enough we get
\begin{equation}                                      \label{eq:3.34}
|v_1(x,t,\omega)-v_2(x,t,\theta)|^2=4\sin^2\frac{\alpha}{2}+O(\e).
\end{equation}

Thus the probability density (\ref{eq:3.34})  depends on the magnetic 
flux $\alpha$.  
Therefore the magnetic potentials  belonging to  different gauge  equivalence
classes  make different  physical  impact.  This proves the magnetic AB effect.

\subsection{The three-dimensional case}

The constructions of the subsection 3.1 can be carried out  in the case of three 
dimensions.  Consider,  for example, a toroid  $\Omega_1$  in $\R^3$  as in 
Tonomura et al experiment (cf [T et al]).  Let $x^{(0)}$  be a point outside
 of $\Omega_1$
and let $\gamma_1=\{x=x^{(0)}+s\omega,\ s\leq 0\}$
be a ray passing  through the hole of the toroid.  As in subsection 3.1  we can construct a solution $v_1(x,t,\omega)$  of the form (\ref{eq:3.26}).
In the case $n\geq 3$  dimensions there are $(n-1)$  orthogonal unit vectors
$\omega_{\perp j},1\leq j\leq n-1$,  such that $\omega\cdot\omega_{\perp j}=0,\ 
1\leq j\leq n-1$,  and  we have to replace 
$\chi_0(\frac{(x-x^{(0)})\cdot\omega_\perp}{\delta_1})$
in (\ref{eq:3.26})  by
$\Pi_{j=1}^{n-1}\chi_0(\frac{(x-x^{(0)})\cdot\omega_{\perp j}}{\delta_1})$.
Let
$\gamma_2=\{x=x^{(0)}+s\theta,s\leq 0\}$
be a ray  passing  outside  of toroid and
 let  $v_2(x,t,\omega)$  be the corresponding solution  
of the form (\ref{eq:3.27}).  As in subsection 3.1  we get
$$
|v_1(x,t,\omega)-v_2(x,t,\theta)|^2=4\sin^2\frac{\alpha}{2}+O(\e),
$$
where $\alpha=\int_\gamma A(x)\cdot dx,\ \gamma$  is a closed simple curve
encircling $\Omega_1$  and we assume  that  the angle between $\omega$  and $\theta$
is small.  

Assuming that $\alpha\neq 2\pi n,\forall n\in \Z$,  we obtain that the 
probability density $|v_1-v_2|^2$  depends on $\alpha$  and this proves 
the AB effect.




\subsection{The case of several  obstacles}

Let $\Omega_j,\ 1\leq j\leq m,\ m>1,$
be  obstacles in $\R^2$,  and let  $\alpha_j=\frac{e}{hc}\int_{\gamma_j} A(x)\cdot dx$ 
be the magnetic fluxes generated by magnetic fields shielded in $\Omega_j,\ 
1\leq j\leq m$.  Suppose that some $\alpha_j$  satisfy  the condition (\ref{eq:3.30}).
If the obstacles are close to each other  it 
is impossible to repeat the construction of the subsection 3.1 separately 
for each  $\Omega_j$.  Note  that if the total flux 
$\sum_{j=1}^m\alpha_j=2\pi p,\ p\in\Z$,  then  treating $\Omega=\cup_{j=1}^m\Omega_j$ 
as one obstacle  we will miss the magnetic AB effect.

In this subsection we show how to determine all 
$\alpha_j (\mbox{mod\ }2\pi p),\ 1\leq j\leq m,$
using  the broken rays.

We shall  introduce some notations.

Let  $x^{(1)}\not\in \Omega=\cup_{j=1}^m\Omega_j$.  Denote by 
$\gamma=\gamma_1\cup\gamma_2\cup...\cup\gamma_r$  the broken ray starting 
at $x^{(1)}$  and reflecting at $\Omega$  at points $x^{(2)},...,x^{(r)}$.
The last  leg $\gamma_r$  can be extended to the infinity.
Denote  by $\omega_p,\ 1\leq p\leq r$,  the directions of $\gamma_p$.
The equations of $\gamma_1,...,\gamma_r$  are $x=x^{(1)}+s\omega_1,\ 
s_1=0\leq s\leq s_2,\ x=x^{(2)}+s\omega_2,
\ s_2\leq s\leq s_3,...,x=x^{(r)}+s\omega_r,\ s_r\leq s<+\infty$.
Here $s_p$  are  such that $x(s_p)=x^{(p)},\ 1\leq p\leq r$.
Denote by 
$\tilde\gamma=\tilde\gamma_1\cup\tilde\gamma_2\cup...\cup\tilde\gamma_r$ 
the lifting of $\gamma$  to $\R^2\times(0,+\infty)$,  where
the equations of $\tilde\gamma_p$  are  
$x=x^{(p)}+s\omega_p,\ t=s,\ s_p\leq s\leq s_{p+1},\ s_{r+1}=+\infty$.
Note  that the times when $\tilde\gamma$  hits the obstacles are 
$t_p=s_p,\ 2\leq p\leq r$.

Let $V_0$ be a small neighborhood 
 of $x^{(1)}$.  Denote by  $\gamma_y=\cup_{p=1}^r\gamma_{py}$  the broken ray that starts
at  $y\in V_0$ at $t=0$. 
 We assume  that $\gamma_{1y}$  has  the form $x=y+s\omega_1,\ 0\leq s\leq s_2(y),$
where  $x^{(2)}(y)=y+s_2(y)\omega_1$  is the point where
$\gamma_{1y}$  hits  $\partial\Omega$.  In particular,
$\gamma_{x^{(1)}}=\gamma$.  Let  $U_0(t)=\{x=x(t)\}$  be the set  of endpoints 
at the time $t$  of $\tilde\gamma_y,\ y\in V_0$.  Note that there is a
one-to-one correspondence between $y\in V_0$  and $x(t)\in U_0(t)$.
Therefore  we shall denote the broken ray starting  at $y\in V_0$  and
ending at  $x(t)$  at the time $t$ by  $\gamma(x(t))$  instead of $\gamma_y$.
As in [E3], [E4]  we can construct a geometric optics solution of
$(\frac{h^2}{2m}\frac{\partial^2}{\partial t^2}+H)w_N=0$  in
$(\R^2\setminus\Omega)\times(0,+\infty)$  in the form
\begin{multline}                                   \label{eq:3.35}
w_N(x,t)
\\
=\sum_{p=1}^r\sum_{n=0}^Ne^{i\frac{mk}{h}(\psi_p(x)-t)}\frac{a_{pn}(x,t)}{(ik)^n}
+
\sum_{p=1}^r\sum_{n=0}^Ne^{i\frac{mk}{h}(\psi_p(x)+t)}\frac{b_{pn}(x,t)}{(ik)^n},
\end{multline}
where
\begin{align}                                     \label{eq:3.36}
|\nabla \psi_p(x)|=1,
&\ \ \frac{\partial\psi_p(x^{(p)})}{\partial x}=\omega_p,\ \ 1\leq p\leq r,
\\
\nonumber
&
\psi_1(x)=x\cdot\omega_1.
\end{align}
We have that $a_{pn}(x,t)=b_{pn}(x,-t)$
and $a_{pn}(x,t)$ satisfy the transport equations
\begin{multline}                                \label{eq:3.37}
\frac{\partial a_{pn}}{\partial t}+\frac{\partial\psi_p(x)}{\partial x}
\cdot\frac{\partial a_{pn}}{\partial x}+\frac{1}{2}\Delta\psi_p a_{pn}
-i\frac{e}{hc}A(x)\cdot\frac{\partial\psi_p}{\partial x}a_{pn}=f_{pn}(x,t),
\\
1\leq p\leq r,\ 0\leq n\leq N,\qquad\qquad
\end{multline}
where
$f_{p0}=0,\ f_{pn}$  depend on $a_{pj}$  for $n\geq 1,\ 0\leq j\leq n-1$.
The following boundary conditions hold on $\partial\Omega\times(0,+\infty)$:
\begin{align}                                  \label{eq:3.38}
&\psi_p\big|_{\partial\Omega\times(0,+\infty)}
=\psi_{p+1}\big|_{\partial\Omega\times(0,+\infty)},\ \ 1\leq p\leq r-1,
\\
\nonumber
&a_{pn}\big|_{\partial\Omega\times(0,+\infty)}=
-a_{p+1,n}\big|_{\partial\Omega\times(0,+\infty)},\ \  1\leq p\leq r-1.
\end{align}
Conditions (\ref{eq:3.38}) imply  that
$$
w_N\Big|_{\partial\Omega\times(0,+\infty)}=0.
$$
We  impose the following initial conditions:
\begin{align}                                 \label{eq:3.39}
&a_{10}(x,0)=\frac{1}{2}\chi_0\Big(\frac{(x-x^{(1)})\cdot\omega_{1\perp}}{\delta_1}\Big)
\chi_0\Big(\frac{(x-x^{(1)})\cdot\omega}{\delta_2}\Big),
\\
\nonumber
&a_{1n}(x,0)=0,\ \ n\geq 1.
\end{align}
We assume that 
$\delta_1,\delta_2$  in (\ref{eq:3.37}) are small,  so that the support of
 the first sum  in (\ref{eq:3.35})  is contained in a small 
neighborhood
of $\tilde\gamma=\cup_{p=1}^r\tilde\gamma_p$.  
We define $a_{pn}(x,t)$  as zero  outside of this neighborhood of $\tilde\gamma$.

Let $x^{(0)}\in\gamma_r$  and  $(x^{(0)},t^{(0)})$ be a corresponding  point  on
$\tilde\gamma_r$.
It was shown in [E3], [E4] that
\begin{equation}                                    \label{eq:3.40}
a_{r0}(x,t)=c_0(x,t)\exp\Big(\frac{ie}{hc}\int_{\gamma(x,t)}A(x)\cdot dx\Big) 
+O\Big(\frac{1}{k}\Big),
\end{equation}
where
$\gamma(x,t)$  is the broken  ray  starting  in a neighborhood
of $x^{(1)}$  at $t=0$  and ending  at $(x,t),\ c(x,t)\neq 0$
on $\gamma(x,t)$.

As in subsection 3.1  we have that $a_{rn}(x,t),n\geq 1$
satisfy  the estimates  of the form  (\ref{eq:3.5}).

Let  $\tilde\gamma(x^{(0)},t^{(0)})$  be  the broken ray starting  at
$(x^{(1)},0)$  and ending  at $(x^{(0)},t^{(0)})$,  where  $x^{(0)}\in \gamma_r$.
Let
\begin{equation}                                        \label{eq:3.41}
u_N(x,t)=\frac{e^{-i\frac{\pi}{4}}\sqrt m}{\sqrt{2\pi ht}}
\int_{-\infty}^\infty e^{\frac{imx_0^2}{2ht}}w_N(x,x_0)dx_0
\end{equation}
where $w_N(x,x_0)$  is  the same as in (\ref{eq:3.35}).
We assume in this subsection  that  
\begin{equation}                                      \label{eq:3.42}
t=\frac{t'}{k},\ \ 0\leq t'\leq T'. 
 \end{equation}
Applying the stationary phase method  to (\ref{eq:3.41})
and using (\ref{eq:3.40}), (\ref{eq:3.42})  we get for $x$  belonging to
a neighborhood of $x^{(0)}$
\begin{multline}                                    \label{eq:3.43}
u_{N}(x,t)
\\ 
=\exp\Big(i\big(-\frac{mk^2t}{2h}+\frac{mk}{h}\psi_r(x)\big)\Big)
c_0(x,kt)\exp\Big(\frac{ie}{hc}\int_{\gamma(x,t')}A(x)\cdot dx\Big)
\\
 +O\Big(\frac{1}{k}\Big),
\end{multline}
where $t'=kt,\ t'$  belongs to a neighborhood of $t^{(0)}$.

Analogously to subsection 3.1 we get  that there exists $R_N(x,t)$
such that $R_N(x,t)=O(\frac{1}{k}),\ t=\frac{t'}{k},\ 0\leq t'\leq T'$,  and
\begin{equation}                                   \label{eq:3.44}
u(x,t)=u_N(x,t)+R_N(x,t)
\end{equation}
is the exact solution of (\ref{eq:1.1})  with the boundary
conditions $u\big|_{\partial\Omega\times(0,\frac{T'}{k})}=0$  and the initial
condition
\begin{equation}                                 \label{eq:3.45}
u(x,0)=\chi_0\Big(\frac{(x-x^{(1)})\cdot\omega_{1\perp}}{\delta_1}\Big)
\chi_0\Big(\frac{(x-x^{(1)})\cdot\omega_1}{\delta_2}\Big).
\end{equation}
\qed

Denote by $\tilde\beta=\{x=x^{(2)}+s\theta,\ t=s,\ 0\leq s\leq t^{(0)}\}$
the ray
starting at $(x^{(2)},0)$  and  ending  exactly at  the point $(x^{(0)},t^{(0)})$.
Analogously to (\ref{eq:3.27})  we can construct a solution $v(x,t)$
of (\ref{eq:1.1}),  satisfying  (\ref{eq:1.2})
and such  that
\begin{multline}
\nonumber
v(x,t)=\chi_0\Big(\frac{(x-x^{(2)})\cdot\theta_{\perp}}{\delta_1}\Big)
\chi_0\Big(\frac{(x-x^{(2)})\cdot\theta-kt}{\delta_2}\Big)c_1(x,kt)
\\
\cdot
\exp\Big(-i\frac{mk^2t}{2h}+i\frac{mk}{h}\theta
+\frac{ie}{hc}\int_0^{t'}A(x^{(2)}+s\theta)\cdot\theta dx\Big)
+O\Big(\frac{1}{k}\Big),
\end{multline}
where 
$t=\frac{t'}{k},\ (x,t')\in U_0$,  where $U_0$  is a neighborhood of
 $(x^{(0)},t^{(0)})$.

We choose initial conditions $c_1(x,0)$ such that  (cf.  (3.43))
$$
c_1(x^{(0)},t^{(0)})=c(x^{(0)},t^{(0)}).
$$
Note that 
$$
\int_0^{t'}\theta\cdot A(x^{(2)}+s\theta)\cdot \theta ds
=\int_{\beta(x,t')}A\cdot dx.
$$
As in (\ref{eq:3.28}) we have  near $(x^{(0)},\frac{t^{(0)}}{k})$
$$
|u(x,t)-v(x,t)|^2=|c(x^{(0)},t^{(0)})|^24\sin^2\frac{1}{2}
\Big(\frac{mk}{h}(\psi_r(x)-\theta\cdot x)+I_1-I_2\Big)+O(\e),
$$
where 
$$
I_1=\frac{e}{hc}\int_{\gamma(x^{(0)},t^{(0)})} A\cdot dx,\ \ \ \
I_2=\frac{e}{hc}\int_{\beta(x^{(0)},t^{(0)})} A\cdot dx.
$$
Choose  $k_n>k_0$ such  that
$$
\frac{mk_n}{h}(\psi_r(x^{(0)},t^{(0)})-x^{(0)}\cdot\theta)=2\pi n,\ n\in \Z,
$$
and choose the initial points $x^{(1)}$ an $\gamma_1$  and $x^{(2)}$  on $\beta$
far enough from $\Omega$ to have the integral 
$$
I_3=\frac{e}{hc}\int_\sigma A\cdot dx
$$
small.  Here $\sigma$  is the straight  line connecting   $x^{(1)}$  and  $x^{(2)}$
and not  intersecting $\Omega$.  Then if the neighborhood $U_0$
is small enough we get
$$
|u(x,t)-v(x,t)|^2=
|c(x^{(0)},t^{(0)})|4\sin^2\frac{\alpha}{2}+O(\e),
$$
where  $\alpha=I_1-I_2+I_3,\ t=\frac{t'}{k},\ (x^{(0)},t')\in U_0$.
Note that $\alpha$  is the sum of magnetic fluxes of obstacles that are bounded  
by  $\gamma\cup (-\beta)\cup\sigma$.  Varying  $\gamma$ and $\beta$  at least
$m$ times we get enough linear relations to recover all 
$\alpha_j(\mbox{mod\ }2\pi n),\ 1\leq j\leq m$.

{\bf Remark 3.1.}
Even in the case of one  obstacle  it is sometimes convenient 
to consider broken rays reflecting from the artificial boundaries (mirrors).
Note that mirrors were used in the original AB experiment.

\section{The proof  of the electric Aharonov-Bohm effect}
\label{section 4}
\init

Let $D$  be a domain in $\R^n\times[0,T]$ and let $D_{t_0}=
D\cap\{t=t_0\}$.  Assume that $D_{t_0}$  depends continuously on
$t_0\in [0,T]$  and that normals to $D\setminus(\overline{D_0}\cup
\overline{D_T})$
 are not parallel to the $t$-axis.
Suppose that  the magnetic potential $A(x,t)=0$  in $D$  and
consider the Schr\"odinger equation in $D$:
\begin{equation}                                     \label{eq:4.1}
ih\frac{\partial u(x,t)}{\partial t}+\frac{h^2}{2m}\Delta u(x,t)- e V(x,t)u(x,t)=0,\ \ 
0<t<T,
\end{equation}             
with zero Dirichlet boundary condition
\begin{equation}                             \label{eq:4.2}
u\big|_{\partial D_t}=0\ \ \mbox{for}\ \ 0<t<T
\end{equation}
and nonzero initial condition
\begin{equation}                             \label{eq:4.3}
u(x,0)=u_0(x),\ \ x\in D_0.
\end{equation}
Suppose that electric field $E=\frac{\partial V}{\partial x}=0$  in $D$.
If $D_t$  is connected for all  $t\in [0,T]$  then 
$V(x,t)=V(t),$  i.e.  $V(t)$  does not depend  on $x$.  Making  the gauge transformation
\begin{equation}                                \label{eq:4.4}
v(x,t)=\exp\Big(i\frac{e}{h}\int_0^t V(t')dt'\Big)u(x,t)
\end{equation}
we get  that $v(x,t)$  satisfies the Schr\"odinger equation
\begin{equation}                                 \label{eq:4.5}
ih\frac{\partial v}{\partial t}+\frac{h^2}{2m}\Delta v(x,t)=0,
\end{equation}
where 
\begin{equation}                                  \label{eq:4.6}
v\big|_{\partial D_t}=0\ \ \mbox{for}\ \ 0<t<T,
\end{equation}
\begin{equation}                                  \label{eq:4.7}
v(x,0)=u_0(x),\ x\in D_0.
\end{equation}Therefore $V(t)$  is gauge equivalent  to zero electric potential, 
 i.e. there is no electric AB effect in the case when $D_{t_0}$  are connected 
for all $t_0\in [0,T]$.  To have the electric AB effect the domain $D$ 
 must have a more complicated topology.
\qed

We shall describe an electric AB  effect when $A=0, E=0$  
in $D$  but the electric potential $V(x,t)$  is not gauge equivalent to the zero potential.

Consider the cylinder $D^{(1)}=\Omega\times[-T-1,T+1]$  where $\Omega$ is the unit disk 
$x_1^2+x_2^2<1$.  Let $D_{t_0}^{(1)}=D^{(1)}\cap\{t=t_0\}$.  For any $t_0\in [-T,T]$  
denote  by
$D_{t_0}^{(2)}$  the part of $D_{t_0}^{(1)}$  
where $|x_1|<\frac{1}{2}-\frac{1}{2}\frac{t_0^2}{T^2}$.
Let $D^{(2)}$  be the union of $D_{t_0}^{(2)}$  when $-T\leq t_0\leq T$  and let
$D=D^{(1)}\setminus\overline D^{(2)}$.  The domains $D_{t_0}$ change the topology when $t$   
changes on $[-T-1,T+1]$  (see Fig. 2).

 The slices  $D_{t_0}=D\cap\{t=t_0\}$ are disks when $t_0\in[-1-T,-T)$
and $t_0\in (T,T+1]$ and they are not connected domains 
$D_{t_0}=D_{t_0}^{(1)}\setminus D_{t_0}^{(2)}$
when $t_0\in(-T,T)$.  If $\gamma$  is a closed contour surrounding $D^{(2)}$ 
 then $\gamma$  is
not homotopic to a point.

We consider the Schr\"odinger  equation
(\ref{eq:4.1})  in $D$
with nonzero initial condition
(\ref{eq:4.3})                             
for $t=-T-1$
and the zero boundary  condition (\ref{eq:4.2})
for $-T-1<t\leq T+1$.

   The gauge  group  $G(\overline D)$  consists  of all  $g(x,t)$  in $\overline D$   
such that $|g(x,t)|=1$.
It follows from the topology  of $D$  that any $g(x,t)\in G(\overline D)$  has the form:
$$
g(x,t)=e^{i n\theta+\frac{i}{h}\varphi(x,t)},
$$
where $\varphi(x,t)$  is 
real-valued and
differentiable in $\overline D$  and $\theta$  is the polar angle  in
$(x_1,t)$-plane.  If $V_1(x,t)$ and $V_2(x,t)$  are gauge equivalent  and if $\gamma$  is
a closed contour in $D$ encircling $D^{(2)}$,   then
$$
e\int_\gamma V_1(x,t)dt -e\int_\gamma V_2(x,t)dt
=ih\int_\gamma g^{-1}(x,t)\frac{\partial g}{\partial t}dt
=2\pi h n,\ \ n\in \Z.
$$
Suppose $V(x,t)=0$  for $t\in [-T-1,-T+\e]$  and $[T-\e,T+1]$,  where $\e>0$,  and 
suppose $V(x,t)=V_1(t)$  when 
$t\in[-T+\e,T-\e],\ x_1>0,\ V(x,t)=V_2(t)$  when $t\in[-T+\e,T-\e],\ x_1<0$.   Then
$E=\frac{\partial V}{\partial x}=0$  in $D$.
\\
\\
\begin{tikzpicture}
\draw  (-6,0)  arc(180:360:  6 and 2); 
\draw[dashed] (6,0) arc(0:180:6 and 2);
 
\draw  (-6,-1.5)  arc(180:360:  6 and 2); 
\draw[dashed] (6,-1.5) arc(0:180:6 and 2);

\draw  (-6,6)  arc(180:360:  6 and 2); 
\draw[dashed] (6,6) arc(0:180:6 and 2);

\draw (0,7.5) ellipse (6 and 2);
\draw (-6,-1.5) -- (-6,7.5);
\draw (6,-1.5) -- (6,7.5); 
 
\draw[dashed] (-4,-1.5) -- (4,1.5);
\draw[dashed] (-4,4.5) -- (4,7.5);

\draw (-4,-1.5) -- (-4,4.5);
\draw[dashed] (4,1.5) --(4,7.5);


\draw (-4,4.5)  arc (30:-30:6 and 6);

\draw (-4,4.5)  arc (150:210:6 and 6);


\draw[dashed] (4,7.5) arc(30:-30: 6 and 6);
\draw[dashed] (4,7.5) arc(150:210: 6 and 6);

\draw[dashed] (-3.2,1.5) -- (4.8,4.5);

\draw[dashed] (-3.3,0.7) -- (4.7,3.7);
\draw[dashed] (-3.4,0) -- (4.6,3);
\draw[dashed] (-3.65,-0.7) -- (4.3,2.2);

\draw[dashed] (-3.65,3.7) -- (4.3,6.7);
\draw[dashed] (-3.4,3) -- (4.6,6);
\draw[dashed] (-3.3,2.2) -- (4.7,5.2);

\draw (7, -1.5) node {$-T-1$};

\draw (6.7, 6) node {$T$};
\draw (6.7, 0) node {$-T$};
\draw (7, 7.5) node {$T+1$};


\end{tikzpicture}  

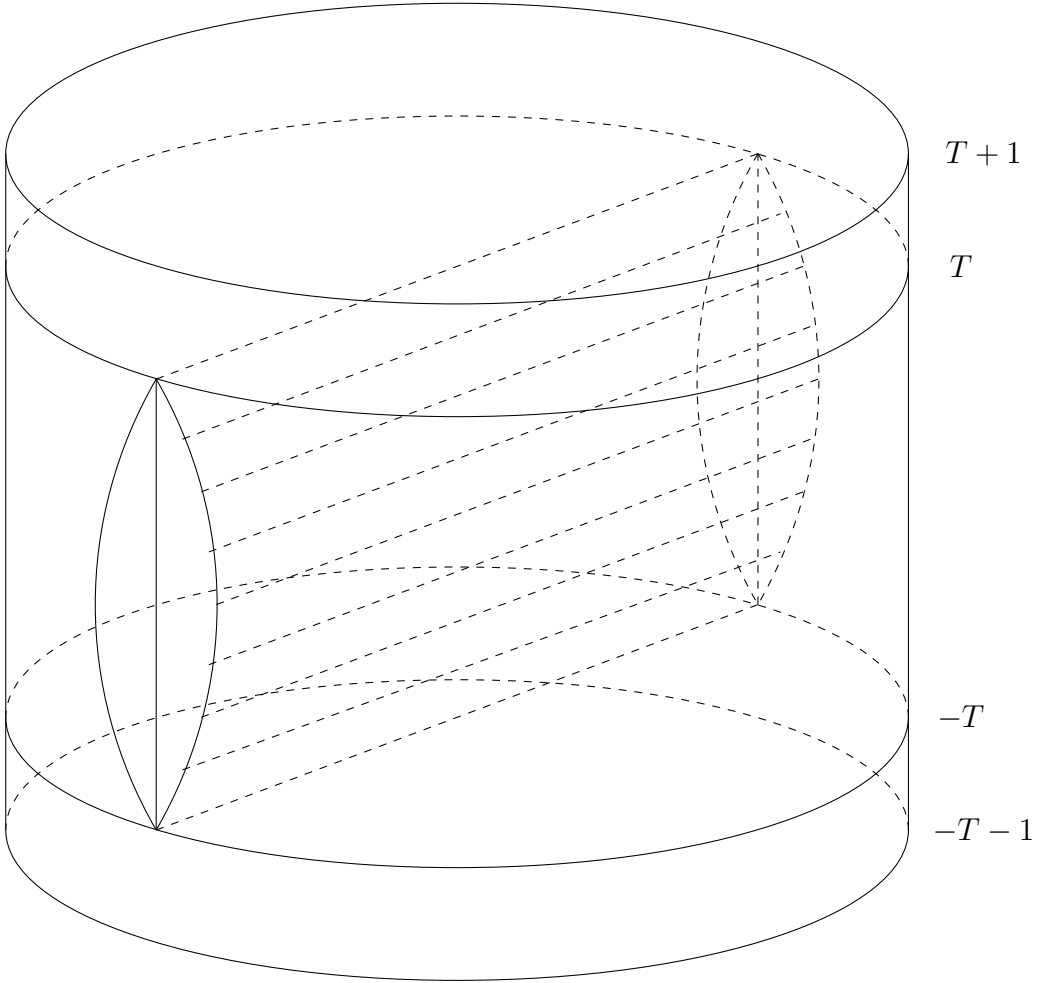
Figure 2.
\\

Denote $\alpha_j=\frac{e}{h} \int_{-T+\e}^{T-\e} V_j(t)dt,\ j=1,2,$  and 
suppose $\alpha_1-\alpha_2\neq 2\pi  n, \ \forall n\in\Z$.
Since the electric flux $\alpha=\frac{e}{h}\int_\gamma V(x,t)dt=\alpha_1-\alpha_2\neq 2\pi n,
\ \forall n$  the electric potential $V(x,t)$  is not gauge  equivalent to the zero  potential.                              

Let $u(x,t)$  be the solution of (\ref{eq:4.1}) 
in $D$
with the initial and boundary conditions 
(\ref{eq:4.2}),  (\ref{eq:4.3})  and let  $v(x,t)$  
be  the solution of (\ref{eq:4.5})  in $D$
with $V(x,t)=0$ and the same initial and boundary conditions as $u(x,t)$.

We shall show that if $\alpha=\frac{e}{h}\int_\gamma V(x,t)dt=\alpha_1-\alpha_2\neq 2\pi m,\forall m\in\Z$
then the probability  densities $|u(x,t)|^2$  and $|v(x,t)|^2$
differ near $t=T$.  Therefore  $V(x,t)$  makes a physical impact  
different  from the impact of the zero potential.
This will prove  the electric AB effect.
Since $u(x,t)$  and $v(x,t)$  have the same initial and boundary conditions (\ref{eq:4.2}),
(\ref{eq:4.3})  and $V(x,t)=0$  for $t\in[-T-1,-T+\e]$  we have that 
$u(x,t)=v(x,t)$  for $t\leq -T+\e$.   Denote  by $\Pi_1$  and $\Pi_2$
two connected components of $(D^{(1)}\setminus D^{(2)})\cap(-T+\e,T-\e)$.  Note that
in $\Pi_i,i=1,2,$  we have 
$$
u(x,t)=v(x,t)\exp\left(-i\frac{e}{h}\int_{-T+\e}^tV_i(t')dt'\right),\ \ (x,t)\in\Pi_i.
$$
Let 
 $c(x,t)=1$  for  $t\leq -T+\e,\ c(x,t)=e^{-i\frac{e}{h}\int_{-T+\e}^tV_i(t')dt'}$  
in $\Pi_i$ 
for $t\in (-T+\e,T-\e),i=1,2.$  Then
$u(x,t)=c(x,t)v(x,t)$,  i.e.  $V(x,t)$  is gauge equivalent  to zero in 
$D\cap(-T-1,T-\e)$.  However,  $\lim_{t\rightarrow T}c(x,t)$  is equal to
$\exp(-i\frac{e}{h}\int_{-T+\e}^{T}V_1(t')dt')$  for  $x_1>0$  and is equal to
$\exp(-i\frac{e}{h}\int_{-T+\e}^{T}V_2(t')dt')$ when $x_1<0$.
Since 
$\exp(-i\frac{e}{h}\int_{-T+\e}^TV_1(t')dt')
\neq \exp(-i\frac{e}{h}\int_{-T+\e}^TV_2(t')dt')$
we have that $c(x,t)$  is discontinuous at $x_1=0,t=T$.
\qed

Note that $u(x,t)$  and $v(x,t)$  satisfy the same equation (\ref{eq:4.5}) in 
$D\cap\{T\leq t\leq T+1\}$
and 
$$
u(x,T)=e^{i\alpha_1}v(x,T)\ \ \mbox{for}\ \ x_1>0,
$$
$$
u(x,T)=e^{i\alpha_2}v(x,T)\ \ \mbox{for}\ \ x_1<0.
$$
We shall show  that the probability densities $|u(x,t)|^2$  and $|v(x,t)|^2$
are not equal identically for $T< t<T+\e$.  Therefore the physical impact 
of the electric potential $V(x,t)$  differs from
the impact  of the zero potential.

To prove that $|u(x,t)|^2\not\equiv |v(x,t)|^2$  in $D$  we consider 
$w(x,t)=e^{-i\alpha_2}u(x,t)$.   Then $w(x,T)=v(x,T)$  for  
$x_1<0,w(x,T)=e^{i(\alpha_1-\alpha_2)}v(x,T)$ for $x_1>0$.
\begin{pro}                                \label{prop:4.1}
If $|w(x,t)|^2=|v(x,t)|^2$  for $T\leq t\leq T+\e$  and $v(x,T)=w(x,T)$  for $x_1<0$,
then $v(x,t)=w(x,t)$  for all $T< t< T+\e,\ x_1^2+x_2^2\leq 1$.
\end{pro}

{\bf Proof of Proposition \ref{prop:4.1}}.
Let $v(x,t)\neq 0$ in some neighborhood $O$  of $(x_1^{(0)},0,T),x_1^{(0)}<0$.  Denote 
$R(x,t)=|v(x,t)|,\ \Phi(x,t)=\arg v(x,t)$,  i.e.
$v(x,t)=R(x,t)e^{i\Phi(x,t)}$.   Substituting in (\ref{eq:4.5})  and separating
the real and the imaginary parts we get
\begin{equation}                             \label{eq:4.8}
-hR_t=\frac{h^2}{2m}(2\nabla R\cdot\nabla \Phi+R\Delta\Phi),
\end{equation}
\begin{equation}                               \label{eq:4.9}
h\Phi_t R=\frac{h^2}{2m}(\Delta R-R|\nabla\Phi|^2).
\end{equation}
Suppose $R$  is given.  Then  (\ref{eq:4.9})  is a first order 
partial differential
equation in $\Phi$  and 
therefore
the initial data $\Phi(x,T)$  near  $(x_1^{(0)},0)$  uniquely  determines $\Phi(x,t)$  
in the neighborhood $O$.  
Let $w=R_1(x,t)e^{i\Phi_1(x,t)}$.
Note that $w(x,t)$ also satisfies  (\ref{eq:4.5})  
for  $t>T$  and $\Phi_1$  satisfies (\ref{eq:4.9}).
Since $R=R_1$  in $O$  and, since $\Phi_1(x_1,0,T)=\Phi(x_1,0,T)$  in $O$, we have that 
$w(x,t)=v(x,t)$  in $O$.
Then $w(x,t)=v(x,t)$  for $T< t< T+\e,\ x_1^2+x_2^2<1,$  
by the unique continuation property (see [I], section 6). 
By the continuity in $t$  we get  $v(x,T)=w(x,T)$  for $x_1>0$
and  this is a contradiction with 
$w(x,T)=e^{i(\alpha_1-\alpha_2)}v(x,T)$  for  $x_1>0$.  Therefore 
$|u(x,t)|^2\not\equiv |v(x,t)|^2$  for  $T<t<T+\e$, i.e. the AB effect holds.
\qed

{\bf Remark 4.1.}  In the proof of Proposition \ref{prop:4.1} we used that 
$R(x,t)=|v(x,t)|\not \equiv 0$  for
$t=T$ and $x_1<0$  and $|v(x,t)|\not\equiv 0$  for  $t=T$  and $x_1>0$.
We  shall  show that  this can be achieved by the appropriate choice  of
the initial condition $u_0(x)$  in (\ref{eq:4.7}).  Choose any $v(x,T)$  such 
that $v(x,T)\not\equiv 0$  for $x_1>0$  and  $v(x,T)\not\equiv 0$  for $x_1<0$.

Solve  the backward initial value  problem  for (\ref{eq:4.5})
with the boundary condition (\ref{eq:4.6}) and the initial  condition $v(x,T)$  for 
$t=T$.  Then  we take  $v(x,-T-1)$  as the initial condition $u_0(x)$
in (\ref{eq:4.7})  and 
(\ref{eq:4.3}).

\subsection{A possible physical experiment  to demonstrate the electric
AB effect}

The class of domains $D$  with nontrivial topology  that leads
to the electric AB effect is large. Below  we give an example that 
can lead to 
a physical experiment  to demonstrate the electric AB effect.

Denote  by $\Omega(\tau)$  the interior  of 
the unit disk  $x_1^2+x_2^2\leq 1$ 
with removed 
 two  
 parts $\Delta(\tau)$  and  $\Delta(-\tau)$  depending on the parameter
$\tau,\ 0\leq \tau\leq \frac{1}{2}$
 (see Fig. 3).

\begin{tikzpicture}
\draw (-1.5,0) -- (1.5,0);
\draw (0,-1.5) -- (0,1.5);
\draw (0,0) circle (1cm);
\draw (-0.9539,0.3) -- (-0.25,0.3);
\draw (-0.9539,-0.3) -- (-0.25,-0.3);
\draw (-0.25, 0.3)-- (-0.25, -0.3);
\draw (-0.54, 0.12) node {$-\tau$};

\draw (0.9539,0.3) -- (0.25,0.3);
\draw (0.9539,-0.3) -- (0.25,-0.3);
\draw (0.25, 0.3)-- (0.25, -0.3);
\draw (0.40, 0.12) node {$\tau$};

\end{tikzpicture}
Fig.3

Let $D$  be  the following domain in $\R^2\times[0,T+1]$:
$$
D_t=\Omega(\frac{1}{2}-t)\ \ \mbox{for}\ \ 0\leq t\leq \frac{1}{2},
$$
$$
D_t=\Omega(0)\ \ \mbox{for}\ \ \frac{1}{2}\leq t\leq T+\frac{1}{2},
$$
$$
D_t=\Omega(t-\frac{1}{2}-T)\ \ \mbox{for}\ \ T+\frac{1}{2}\leq t\leq T+1.
$$
Here $D_{t_0}=D\cap\{t=t_0\}$.

Therefore $\Delta(\tau)$  and  $\Delta(-\tau)$  increase  in size  from
$\tau=\frac{1}{2}$  to $\tau=0$  when  $0\leq t\leq \frac{1}{2}$.  Then they do not move  for  $\frac{1}{2}\leq t\leq T+\frac{1}{2}$.  Note  that $D_t$  consists of
the components $D_t^+$  and $D_t^-$  for 
$\frac{1}{2}\leq t\leq \frac{1}{2}+T$.  When 
$T+\frac{1}{2}\leq t\leq T+1$  the parts $\Delta(\tau)$  and $\Delta(-\tau)$  
return  back to
the initial position $\tau=\frac{1}{2}$. 

Such moving domain is easy to realize experimentally.  We can arrange  that
$V(t)=0$  in $D$  for  $0\leq t\leq \frac{1}{2}+\e$  and for
$\frac{1}{2}+T-\e<t\leq T+1, \ V(t)=V_1(t)$  in $D_t^+, \
V(t)=V_2(t)$ in $D_t^-$  for $\frac{1}{2}\leq t\leq T+\frac{1}{2}$.
Suppose 
$$
\frac{e}{h}\int_{\frac{1}{2}}^{T+\frac{1}{2}}V_1(t)dt-
\frac{e}{h}\int_{\frac{1}{2}}^{T+\frac{1}{2}}V_2(t)dt\neq 2\pi n,\ \forall n\in \Z.
$$  
Then
one can show theoretically  as in Proposition \ref{prop:4.1} that the electric AB effect
holds for $\frac{1}{2}+T<t<1+T$.  Hopefully this can be shown also
experimentally by measuring $|u(x,t)|^2$  for $\frac{1}{2}+T<t<1+T$ and comparing  
these measurements with the measurements for the zero potential in $D$.

We assume that the initial data $u_0(x)$ for $V(t)$ and $V\equiv 0$  are the same.

\section{Combined electric and magnetic AB effect}
\label{section 5}
\init
\subsection{Class of domains}
Consider
the following domain $D\subset\R^n\times(0,T)$:
Let
$T_0=0<T_1<...<T_r=T$.
Denote by $D_{t_0}$ the intersection of $D$  with the plane $t=t_0$.
Then for $t_0\in (T_{p-1},T_p),\ p=1,...,r$,  we have
$D_{t_0}=\Omega_0\setminus\overline{\Omega_p(t_0)}$,  where $\Omega_0$
is a simply-connected
domain  in $\R^n,\ \Omega_p(t_0)=\cup_{j=1}^{m_p}\Omega{pj}(t_0),\ 
\overline{\Omega_{pj}(t_0)}\cap\overline{\Omega_{pk}(t_0)}=\emptyset$
for $j\neq k,\ \overline{\Omega_{pj}(t_0)}\subset\Omega_0,\ \Omega_{pj}(t_0)$ are
smooth domains  (obstacles).  Note that 
$m_p$  may be different  for $p=1,...,r$.   We assume that
$\Omega_p(t_0)$  depends smoothly on $t_0\in (T_{p-1},T_p)$.  Also we assume that $D_{t_0}$  depends continuously on $t_0\in [0,T]$.
Note that some obstacles may merge or split when $t_0$  crosses $T_p,p=1,...,r-1$
(see Fig. 4).
\\
\\
\begin{tikzpicture}
\draw  (-6,0)  arc(180:360:  6 and 2); 
\draw[dashed] (6,0) arc(0:180:6 and 2); 
\draw (0,6) ellipse (6 and 2);
\draw (-6,0) -- (-6,6);
\draw (6,0) -- (6,6); 
\draw (1.5,0) ellipse (0.5 and 0.2);
\draw (4.5,0) ellipse (0.5 and 0.2);
\draw (1.5,6) ellipse (0.5 and 0.2);
\draw (4.5,6) ellipse (0.5 and 0.2);
\draw(1,0) -- (2,3);
\draw(5,0) -- (4,3);
\draw(1,6) -- (2,3);
\draw (4,3) -- (5,6);
\draw(2,0) -- (3,2.5);
\draw(4,0) -- (3,2.5);
\draw(3,3.5) -- (2,6);
\draw (3,3.5) -- (4,6);

\draw(3,3) ellipse (1 and 0.2);
\draw[dashed](3,3) ellipse (0.3  and 2.5);
\draw(3.3,0.3) node {$\gamma_5$};

\draw(-1.5,0) ellipse (0.5 and 0.2);
\draw(-4,3) node {$\Omega^{(0)}$};

\draw(0.7,5.2) node {$\Omega^{(1)}$};
\draw(5.2,5.4) node {$\Omega^{(2)}$};
\draw(0.8,0.8) node {$\Omega^{(3)}$};
\draw(5.35,0.43) node {$\Omega^{(4)}$};
\draw (-4.5,6) ellipse (0.5 and 0.2);
\draw (-2,0) -- (-5,6);
\draw(-1,0) -- (-4,6);

\end{tikzpicture}  

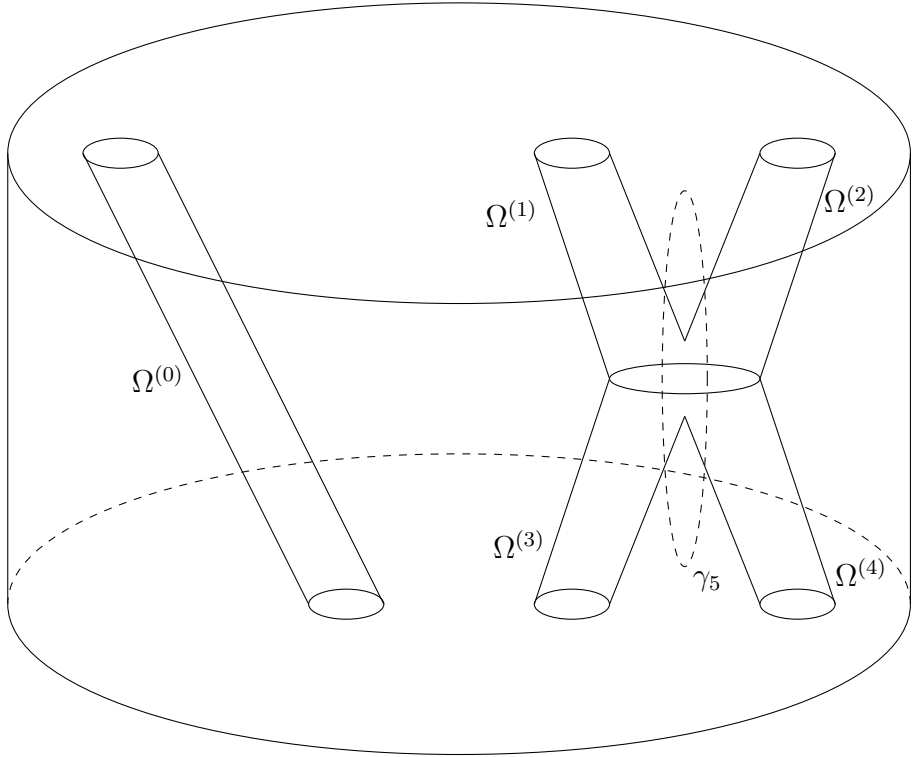
Figure 4.
\\
\

We shall study  the time-dependent Schr\"odinger equation in $D$:
\begin{equation}                                   \label{eq:5.1}
ih\frac{\partial u(x,t)}{\partial t}
-\frac{1}{2m}\sum_{j=1}^n\Big(-ih\frac{\partial}{\partial x_j}
-\frac{e}{c}A_j(x,t)\Big)^2u
-eV(x,t)u(x,t)=0
\end{equation}with smooth time-dependent magnetic potential
$A(x,t)=(A_1(x,t),...,A_n(x,t))$  and electric potential $V(x,t), \ (x,t)\in D$.

We assume that the normals to $\partial\Omega_p(t)$ are not parallel to the $t$-axis
for any $t\in [0,T]$.

The gauge group $G(\overline D)$ is the group of all $C^\infty(\overline D)$ complex-valued
functions $g(x,t)$ such that $|g(x,t)|=1$  in $\overline D$ (cf. \S 1).

Electromagnetic potentials $(A(x,t),V(x,t))$  and $(A'(x,t),V'(x,t))$  are called
gauge equivalent is there exists $g(x,t)\in G(\overline D)$  such that
\begin{eqnarray}                           \label{eq:5.2}
\frac{e}{c}A'(x,t)=\frac {e}{c}A(x,t)+ihg^{-1}(x,t)\frac{\partial g}{\partial x}
\\
\nonumber
eV'(x,t)=eV(x,t)-ihg^{-1}(x,t)\frac{\partial g}{\partial t}.
\end{eqnarray}
We shall consider  the case when 
the magnetic and the electric fields are zero in $D$,
i.e.,  $B=\mbox{curl\ }A(x,t)=0,\ E=-\frac{1}{c}\frac{\partial A(x,t)}{\partial t}-
\frac{\partial V(x,t)}{\partial x}=0,\ (x,t)\in D$.
In this case the integral
\begin{equation}                               \label{eq:5.3}
\alpha=\frac{e}{h}\int_\gamma\frac{1}{c}A(x,t)\cdot dx-V(x,t)dt
\end{equation}
over a closed curve $\gamma$  in $D$  
does not change if we deform  $\gamma$ continuously in $D$. 

The integral (\ref{eq:5.3})  is called the 
electromagnetic flux.  It is easy to describe all gauge
equivalent classes of electromagnetic potentials using the electromagnetic  
fluxes and assuming $B=E=0$ in $D$.

Let $\gamma_1,\gamma_2,...,\gamma_l$  be a basis of the homology group
of $D$,  i.e., any closed contour  $\gamma$  in $D$ is homotopic  to 
a linear  combination of $\gamma_1,...,\gamma_l$  with integer coefficients.
Then fluxes
$$
\alpha_j=\frac{e}{h}\int_{\gamma_j}\frac{1}{c}A(x,t)\cdot dx -V(x,t)dt, \ \ 1\leq j\leq l,
$$
modulo $2\pi n, n\in \Z,$  determine a gauge equivalent class 
of $(A(x,t),V(x,t))$,  i.e.  $(A(x,t),V(x,t))$  and  $(A'(x,t),V'(x,t))$ are 
gauge equivalent  iff 
$\alpha_j-\alpha_j'=2\pi m_j,\ m_j\in \Z,
\ 1\leq j\leq l$,  where $\alpha_j'
=\frac{e}{h}\int_{\gamma_j}\frac{1}{c} A'\cdot dx-V'(x,t)dt$.

In the next section  we shall prove that the electromagnetic potentials 
belonging  to different gauge equivalent classes have a different physical  
impact,  for example,  the probability density $|u(x,t)|^2$  will be different
for some $u(x,t).$

\subsection{The proof of the electromagetic AB effect}
 
We shall introduce localized geometric optics type solutions $u(x,t)$  of 
the Schr\"odinger equation (\ref{eq:5.1})
in $D$ depending on a large parameter $k$ and satisfying the zero initial conditions
\begin{equation}                                  \label{eq:5.4}
u(x,0)=0, \ \ x\in D_0,
\end{equation}
and zero boundary conditions on the boundaries of obstacles
\begin{equation}                              \label{eq:5.5}
u(x,t)\big|_{\partial\Omega}=0,
\end{equation}
where $\Omega\subset\R^n\times(0,T)$  is the union of all obstacles 
$\Omega_p(t),\ t\in [T_{p-1},T_p],\ p=1,...,r$,  and  
$D_0=\Omega_0\setminus\Omega_1(0),\ \Omega_1(0)=\Omega\cap\{t=0\}$.
Such solutions were constructed in [E1].  Suppose $t_0\in (T_{p-1},T_p),\ 1\leq p\leq r$.
Suppose $\gamma(x^{(1)},t_0)=
\beta_1(t_0)\cup...\cup\beta_{d-1}(t_0)\cup\beta_d(x^{(1)},t_0)$  
is a broken ray  in $D_{t_0}$  with legs  $\beta_1(t_0),...,\beta_d(x^{(1)},t_0)$
reflecting at $\partial\Omega_p(t_0)$,  starting  at 
point
$x^{(0)}\in \partial\Omega_0$  and ending at  $x^{(1)}\in D_{t_0}$.

As in [E1]  we can construct an asymptotic solution as $k\rightarrow\infty$ of the form
(cf subsection 3.3):
\begin{equation}                                     \label{eq:5.9}
u_N(x,t)=\sum_{j=1}^d
e^{-\frac{imk^2}{2h}t+i\frac{mk}{h}\psi_j(x,t)}
\ \sum_{n=0}^N\frac{a_{nj}(x,t,\omega)}{(ik)^n},
\end{equation}
where
$\psi_1(x,t)=x\cdot\omega$  and $\mbox{supp\ }u_N(x,t,\omega)$ is contained in a
small neighborhood of $x=\gamma(x^{(1)},t_0),t=t_0$
(see [E1]  for the details).
As it was shown in [E1]  one can find $u^{(N)}(x,t)$  such that $Lu^{(N)}=
-Lu_N=O(\frac{1}{k^{N+1}})$  in $D$,
 $u^{(N)}\big|_{t=0}=0,$  and $u^{(N)}\big|_{\partial\Omega}=0,\ 
u^{(N)}\big|_{\partial\Omega_0\times(0,T)}=0$
and such that $u^{(N)}=O(\frac{1}{k^{N-2}})$. 
Here  $L$  is the right hand side  of (\ref{eq:5.1}). 
 Then 
\begin{equation}                               \label{eq:5.11}
u=u_N+u^{(N)}
\end{equation}is the exact  solution of $Lu=0$  in $D$,
 $u\big|_{t=0}=0,\ x\in D_0$,  $u\big|_{\partial\Omega}=0$ 
for all $0<t<T.$

Let  $t_0\in (T_p,T_{p+1})$  and  let $m_p$  be  the number of the obstacles  in 
$D_{t_0}$. 
It was proven  in [E1], [E3]  that $u(x,t)$  has  the following  form in
the neighborhood $U_0$  of $(x^{(1)},t_0)$:
\begin{equation}                                \label{eq:5.12}
u(x,t)=c(x,t)\exp\Big(-i\frac{mk^2t}{2h}+i\frac{mk}{h}\psi_d(x,t)
+\frac{ie}{hc}\int_{\gamma(x,t)}A(x,t)\cdot dx\Big)+O\Big(\frac{1}{k}\Big),
\end{equation}
Here $c(x^{(1)},t_0)\neq 0$
and $\gamma(x,t)$  is a broken ray in $D_t$ that starts at $(y,t)$,  $(y,t)$ is
close to $(x^{(0)},t_0)$,  and such that the first leg of $\gamma(x,t)$  has the
same direction  as $\beta_1(t_0)$.

Note the difference between asymptotic solution (\ref{eq:3.35})
for the wave equations and asymptotic solution (\ref{eq:5.12})  for the Schr\"odinger 
equations. 
Solution (\ref{eq:3.35}) corresponds to  the broken ray  
$\tilde\gamma=\cup_{j=1}^r \tilde\gamma_j$  in  $\R^2\times(0,+\infty)$
and  solution (\ref{eq:5.12})  corresponds  to the broken ray $\cup_{j=1}^d\beta_j$
in the plane $t=t_0$.
\qed

Let $\gamma_1$  be the ray  $x=x^{(0)}+s\theta,\ s\geq 0,\ t=t_0,$  starting at
$(x^{(0)},t_0)$  and  ending at  $(x^{(1)},t_0)$.   Choose  $x^{(1)}\in \Omega_0$
 such that  $\gamma_1$  does not intersect  $\Omega(t_0)$.
We assume  that $\Omega_0$  is large enough that  such $x^{(1)}$ exists (see Fig.5):  
\\
\\
\begin{tikzpicture}

\draw(-3,5) -- (7,4);

\draw (7,4) --(8,1);
\draw (-1,0)--(8,1);
\draw(-1,0) -- (-3,5);


\draw(7.7,5) circle (1.2);



\draw(9.1,0.5) circle (1.2);

\draw(5.5,2.5) circle (1.4);

\draw(-1.3,-0.3) node {$x^{(0)}$};
\draw(-3,5.3) node {$x^{(1)}$};

\draw(3.5,0.1) node {$\beta_1$};

\draw(7.8,3) node {$\beta_2$};

\draw(3,5) node {$\beta_3$};

\draw(-2.5,3)  node {$\gamma_1$};

\draw(5.5,2) node {$\Omega_1(t_0)$};

\draw(7.7,4.8) node {$\Omega_3(t_0)$};

\draw(9.1,0.2) node {$\Omega_2(t_0)$};

\end{tikzpicture}
\\ 
\\
Fig.5

Let  $v(x,t)$  be a geometric optic solution  
similar to
 (\ref{eq:5.9})
with $d=1$ and 
corresponding to the ray $\gamma_1$. 
  We have,  as in (\ref{eq:5.12}):
\begin{equation}                                \label{eq:5.13}
v(x,t)=c_1(x,t)\exp\Big(-i\frac{mk^2t}{2h}+i\frac{mk}{h}x\cdot\theta
+\frac{ie}{hc}\int_{\gamma_1(x,t)}A(x,t)\cdot dx\Big)+O\Big(\frac{1}{k}\Big).
\end{equation}
We choose the initial value  for  $a_{0}(x,t,\theta)$  (cf.  (\ref{eq:5.9}))
near $(x^{(0)},t_0)$  such that
$$
c_1(x^{(1)},t_0)=c(x^{(1)},t_0).
$$
Consider  $|u(x,t)-v(x,t)|^2$  in a neighborhood 
$\{(x,t): |x-x^{(1)}|\leq\e_0,\ |t-t_0|<\e_0\}$.

As in subsection 3.3 we get for a small neighborhood  of $(x^{(1)},t_0)$
\begin{equation}                                    \label{eq:5.14}
|u(x,t,\omega)-v(x,t,\theta)|^2=
|c(x^{(1)},t_0)|^2\ 4\sin^2\frac{\alpha(t_0)}{2}+O(\e),
\end{equation}
where
$$
\alpha(t_0)=\frac{e}{hc}\Big(\int_{\gamma(x^{(1)},t_0)}
A(x,t_0)\cdot dx-\int_{\gamma_1(x^{(1)},t_0)}A(x,t_0)\cdot dx\Big).
$$
Note that $\alpha(t_0)$  is the sum of the fluxes  of those obstacles
$\Omega_{pj}(t_0),\ 1\leq j\leq m_1$,  that  are encircled by $\gamma\cup\gamma_1$.
As in subsection 3.3, 
varying $\gamma$  and $\gamma_1$  at least  $m_p$  times we can recover (modulo
$2\pi n$)  $\alpha_{pj}(t_0),\ 1\leq j\leq m_p$,  where 
\begin{equation}                                   \label{eq:5.15}
\alpha_{pj}(t_0)
=\frac{e}{hc}\int_{\gamma_{pj}(t_0)}A\cdot dx, \ \ \ 1\leq j\leq m_p,
\end{equation}
and $\gamma_{pj}(t_0)$  is  a simple contour in $D_{t_0}$
encircling $\Omega_{pj}(t_0)$,
$1\leq j\leq m_p$.
Note that $\alpha_{pj}$  are  the same for any  $t_0\in (T_p,T_{p+1})$.
We  can repeat the same arguments for any $t_0\neq T_1,...,T_{p-1}$.

Our class of time-dependent obstacles is such that $D_{t_0}$  is connected for 
any $t_0\in [0,T]$.
It follows from this assumption that a basis of the homology group  of $D$  is
contained  in the set $\gamma_{pj}(t_p),\ 1\leq j\leq m_p,\ t_p\in (T_{p-1},T_{p}),\ 
 1\leq p\leq r$.

Denote such basis by $\gamma^{(1)}(t^{(1)}),...,\gamma^{(l)}(t^{(l)})$.
Then any closed contour $\gamma$  in $D$  is homotopic to a linear combination 
$\sum_{j=1}^ln_j\gamma^{(j)}(t^{(j)})$  where $n_j\in \Z$.  Therefore the flux
\begin{equation}                                      \label{eq:5.16}
\frac{e}{hc}\int_\gamma A\cdot dx-Vdt=\sum_{j=1}^l n_j\alpha^{(j)}(t^{(j)},
\end{equation}
where
$\alpha^{(j)}(t^{(j)})=\frac{e}{hc}\int_{\gamma^{(j)}(t^{(j)})}A\cdot dx$.

Thus
the fluxes $\alpha^{(j)}(t^{(j)}),\ 1\leq j\leq l,\ \mbox{mod\ }2\pi n,\ 
n\in \Z$,  determine the gauge equivalence class of $A(x,t), V(x,t)$.  
Therefore computing the probability densities of appropriate solutions 
we are able  to determine  the gauge equivalence classes of electromagnetic 
potentials.

\subsection{Example 5.1}

Consider the domain shown in Fig. 4.  Let $\gamma_p, \ 0\leq p\leq 4,$
be simple closed curves encircling $\Omega^{(p)}$.
There is also a simple closed curve $\gamma_5$  that is not homotopic 
to any closed curve contained in the plane $t=C$.
Note that $\gamma_1+\gamma_2\approx \gamma_3+\gamma_4$  
where $\approx$  means homotopic.
Also $\gamma_5\approx \gamma_1- \gamma_3$.  Therefore
$\gamma_0,\gamma_1,\gamma_2,\gamma_3$  is a basis of the homology group of $D$.
We made an assumption  that  $D_{t_0}$  is connected  for any  $t_0\in[0,T]$.
 Under this assumption there is always a basis of 
the homology group consisting of "flat"  closed curves,  i.e.  the curves containing
in the planes $t=const$.

Let $\alpha_j$ be the fluxes corresponding  to $\gamma_j$.  
Note  that if $\gamma$  is flat  then  $\alpha=\int_\gamma A\cdot dx$ 
is a magnetic flux. However   
$\alpha_5=\frac{e}{h}\int_{\gamma_5}\frac{1}{c}A\cdot dx-Vdt$
is an electromagnetic flux.
Since $\gamma_5\approx\gamma_1-\gamma_3$  we have that
$\alpha_5=(\alpha_1-\alpha_3)(\mbox{mod\ } 2\pi n),\ n\in \Z$.
\qed

Note that  
our approach allows to calculate first the magnetic fluxes.  To determine
electromagnetic flux we have to represent it as a linear  combination
of magnetic
fluxes as in Example 5.1.

{\bf Remark 5.1.}
In this section we assumed   that  the domains $D_t$  are connected for all
$t\in [0,T]$.
To incorporate  the examples of \S 4  we    have  to allow 
$D_t$  to be not connected  for some $t\in [0,T]$,
i.e.  $D_t=\cup_{k=0}^m D_t^{(0)}$,  where  $D_t^{(0)}$  is the open component
containing the neighborhood of $\partial\Omega_0$  and $D_t^{(k)},\ 1\leq k\leq m$,  are  
other open component that we assume to be simply connected.

Acting as above we can determine the magnetic fluxes $\alpha_j$ in 
$D_t^{(0)} (\mbox{mod\ }2\pi n)$.
Since $D_t^{(k)},\ k\geq 1$,  are  simply connected,  we have that
$\frac{e}{c}A=h\frac{\partial\varphi_k(x,t)}{\partial x}$  in 
$\overline D_t^{(k)}$.  Making the gauge 
transformations  with gauge $e^{i\frac{\varphi_k(x,t)}{h}}$ in $D_t^{(k)}$ we can  
 get that $A=0$  in $D_t^{(k)}$.
Then we   obtain the same situation as in the examples of \S 4.
For example,  in the case of Fig. 4  we can insert a domain of  the form of 
 Fig. 3 inside  the tube $\Omega^{(0)}$.   Note  that in \S 4 we did not determine
the electric flux as in \S 3,  but only find whether the
electric potentials are gauge equivalent or not.

\subsection{Approximation of solutions in $D$ by physically meaningful solutions}

Let  $u(x,t)$ 
be a solution  of (\ref{eq:5.1})  in $(\R^n\times(0,T))\setminus\Omega$,
  where $\Omega\subset\R^n\times(0,T)$
is the union of all
obstacles, $0\leq t\leq T$.  We assume that
\begin{equation}                              \label{eq:5.13}
u\big|_{\partial\Omega}=0
\end{equation}
and 
\begin{equation}                              \label{eq:5.14}
u(x,0)=u_0(x),\ \ x\in\R^n\setminus\Omega(0),
\end{equation}
where $\Omega(t_0)=\Omega\cap\{t=t_0\}$,  in particular, 
$\Omega(0)=\Omega\cap\{t=0\}$.

Initial-boundary  value problem (\ref{eq:5.1}),  (\ref{eq:5.13}), (\ref{eq:5.14})
describes an electron confined to the region
$\R^n\setminus\Omega(t_0),\ 0<t_0<T.$  We shall assume that 
 the initial  data $u_0(x)\in H_2(\R^n\setminus\Omega(0))
\cap \overset{\circ}{H}_1(\R^n\setminus\Omega(0))$  (cf. [E1]). It follows  from [E1]
that  $u(x,t)\in C((0,T),H_2(\R^n\setminus\Omega(t))
\cap \overset{\circ}{H}_1(\R^n\setminus\Omega(t))$,  i.e.  $u(x,t)$  
belongs   
to
 a space of continuous  
functions in $t$  with values in $H_2(\R^n\setminus\Omega(t))
\cap \overset{\circ}{H}_1(\R^n\setminus\Omega(t))$.
Here $u(x,t)\in \overset{\circ}{H}_1(\R^n\setminus\Omega(t))$  
means that $u(x,t)=0$  on $\partial\Omega(t)$.
We shall call 
such solutions $u(x,t)$  physically meaningful. 
In  subsection  5.2  we  considered  solutions $v(x,t)$  of (\ref{eq:5.1})
defined in the domain $D$ only with zero initial conditions
in $\Omega_0\setminus\Omega(0)$, zero boundary  conditions
on $\partial\Omega$ and 
having nonzero values
 on $\partial\Omega_0\times(0,T)$.  We assume that $u(x,t)\in 
C((0,T),H_2(\Omega_0\setminus\Omega(t))$.  Then 
$u\big|_{\partial\Omega_0\times(0,T)}=f$
where $f(x,t)$  is  continuous in $t$,
with values in $H_{\frac{3}{2}}(\partial\Omega_0)$.
We shall denote such class of solutions  by $W(D)$.

We shall show that any solution in $W(D)$  can be approximated by the
restriction to $D$ of
physically meaningful  solutions
$u(x,t)$  such  that $u(x,0)=0$  in $\Omega_0\setminus\Omega(0)$,  
and this will make our proof of electromagnetic  
AB   physically relevant.

Denote by $V$ the Banach space  of functions $u(x,t)$ in $D$  with the norm 
$\|u\|_V=\int_0^T[u]_0dt$,  where
$[v]_0$  is the $L_2$-norm in $D_t=\Omega_0\setminus\Omega(t)$.
Let  $V^*$  be the dual space with the norm  $\|v\|_{V^*}=\sup_{0\leq t\leq T}[v]_0$.
Denote by $K\subset V^*$ the closure in the $V^*$  norm
of the restrictions to $D$ of              
all physically meaningful  solutions such that $v(x,0)=0$  for $x\in D_0
=\Omega_0\setminus\Omega(0)$.

Let $K^\perp$  be the set of all $v\in V$  such that 
$(u,v)=0$  for all $u\in K$.  Here
$(u,v)$  is the extension
of $L_2(D)$  inner product.  Let $f$  be any element  of $K^\perp$.
Extend $f$  by zero   in $(\R^n\setminus\Omega_0)\times (0,T)$.

Let $w(x,t)$ be the solution of 
\begin{eqnarray}                          \label{eq:5.15}
L^*w=f \ \ \mbox{in}\ \ (\R^n\times (0,T))\setminus\Omega,
\\
\nonumber
w\big|_{t=T}=0,\ \ w\big|_{\partial\Omega}=0,
\end{eqnarray}
Note that $w(x,t)\in  C((0,T),\overset{\circ}{H}_1(\R^n\setminus\Omega(t)))$.

By the Green formula  we have
$$
0=(v,f)=(v,L^*w)=ih\int_{\R^n\setminus\Omega(0)}v(x,0)\overline{w}(x,0)dx,
$$
where $v(x,t)$  is a physically meaningful solution.
Since $v(x,0)=0$  in
$\Omega_0\setminus\Omega(0)$ and $v(x,0)$
 is arbitrary in $\R^n\setminus\Omega_0$,  we get that
$$
w(x,0)=0,\ \ x\in \R^n\setminus\Omega_0.
$$

Consider $w(x,t)$  in $(\R^n\setminus\Omega_0)\times(0,T)$.  We assume  that
the electric potential
$V(x,t)=0$  in  $(\R^n\setminus\Omega_0)\times(0,T)$.   If $n=3$  or
$n=2$  and the total magnetic flux 
$\frac{e}{h c}\int_{\partial\Omega_0}A(x,t)\cdot dx=0$,
we can choose the gauge such that $A(x,t)=0$  in $(\R^n\setminus\Omega_0)\times(0,T)$.
In this case  the equation  (\ref{eq:5.1})  has the form 
\begin{equation}                                  \label{eq:5.16}                        
ih\frac{\partial w}{\partial t}+\frac{h^2}{2m}\Delta w=0,
\end{equation}
for $(x,t)\in (\R^n\setminus\Omega_0)\times(0,T)$. 

When $n=2$ and the total  magnetic flux is not zero  we can  choose 
the gauge to make $A(x,t)$ equal to AB potential  in 
$(\R^n\setminus\Omega_0)\times(0,T)$  (cf.  [AB]).
Then  in polar coordinates $(r,\theta)$  we have
in $(R^2\setminus\Omega_0)\times(0,T)$:
\begin{equation}                                            \label{eq:5.17}
ih\frac{\partial w}{\partial t}+\frac{h^2}{2m}\bigg[\frac{\partial^2 w}{\partial r^2}
+\frac{1}{r}\frac{\partial w}{\partial r}
+\frac{1}{r^2}\Big(\frac{\partial}{\partial\theta}+i\alpha\Big)^2
w(r,\theta,t)\bigg]=0.
\end{equation}
\begin{lemma}                                       \label{lma:5.1}
Let $w(x,t)$  be the solution of (\ref{eq:5.1})
 in $(\R^n\setminus\Omega_0)\times(0,T),
w(x,0)=w(x,T)=0,
\ x\in \R^n\setminus\Omega_0,$
and $w(x,t)\in C((0,T),L_2(\R^n\setminus\Omega_0))$.  Then
$w=0$  in $(\R^n\setminus\Omega_0)\times(0,T)$.
\end{lemma}

{\bf Proof:}
Consider  the case of equation (\ref{eq:5.17}).  
The  case of the equation (\ref{eq:5.16})
for $n=2$  or $n=3$ is similar.
Let $R$  be such  that $\overline{\Omega_0}\subset B_R=\{x:|x|<R\}$.  Extend
$w(x,t)$  by zero  for $t>T$ and $t<0$.  Making 
the Fourier transform in $t$  we get for $|x|>R$
\begin{equation}                                        \label{eq:5.18}
-h\xi_0\tilde w(r,\theta,\xi_0)+
\frac{h^2}{2m}\bigg[\frac{\partial^2\tilde w(r,\theta,\xi_0)}{\partial r^2}
+\frac{1}{r}\frac{\partial\tilde w}{\partial r}
+\frac{1}{r^2}\Big(\frac{\partial}{\partial\theta}+i\alpha\Big)^2\tilde w\bigg]=0,
\end{equation}
where
$\theta\in [0,2\pi],\ r=|x|>R$.  Since
$\int_{\R^2\setminus\Omega_0}\int_0^T|w(x,t)|^2dxdt<+\infty$  we have
that $\tilde w(x,\xi_0)$  is continuous  in $\xi_0$  and
$\int_{|x|>R}|\tilde w(x,\xi_0)|^2dx<+\infty$  for  any $\xi_0\in \R$.

The general solution of (\ref{eq:5.18})  in $|x|>R$ has the form (see [AB])
\begin{equation}                                 \label{eq:5.19}
\tilde w(x,\xi_0)=\sum_{n=-\infty}^\infty w_n(r,\xi_0)e^{in\theta},
\end{equation}
where
$$
w_n(r,\xi_0)=a_n(\xi_0)J_{m+\alpha}(kr)+b_n(\xi_0)J_{-m-\alpha}(kr),\ \ 
k=\sqrt{\frac{2m}{h}(-\xi_0)}.
$$
We have
$$
\int_{|x|>R}|\tilde w(x,\lambda)|^2dx=\sum_{n=-\infty}^\infty\int_{r>R}|w_n(r,\xi_0)|^2rdr.
$$
It follows from (\ref{eq:5.19})  
that $\int_{r>R}|w_n(r,\xi_0)|^2dr<+\infty$ for all $\xi_0$  
iff $a_n(\xi_0)=b_n(\xi_0)=0$.  
Therefore  $w(x,t)=0$ for  $|x|>R,\ t\in[0,T]$.
Using the unique continuation property
(cf. [I])  we get that $w(x,t)=0$  in  $(\R^2\setminus\Omega_0)\times(0,T)$.
\qed

\begin{lemma}                                  \label{lma:5.2}
Any $u(x,t)\in W(D)$  can be approximated 
in the $V^*$  norm by physically meaningful solutions $v(x,t)$,  i.e.  
by $v(x,t)\in C((0,T),\ H_2(\R^n\setminus\Omega(t))\cap\overset{\circ}{H}_1(\R^n\setminus\Omega(t))$  that
satisfy  (\ref{eq:5.1})  in $(\R^n\times(0,T))\setminus\Omega$
with the boundary conditions (\ref{eq:5.13}) and the initial conditions (\ref{eq:5.14})
where $u_0(x)=0$  in $\Omega_0\setminus\Omega(0)$.
\end{lemma}

{\bf Proof:}
Let
$u(x,t)\in W(D)$.  We have
$$
(u,f)_D=(u,L^*w)_D,
$$
where  $(u,f)_D$ is the inner product  in $L_2(D)$.  
Note  that $u\big|_{\partial\Omega}=0$
and $w\big|_{\partial\Omega}=0$.  Also $u\big|_{t=0}=0$  and $w\big|_{t=T}=0$
in $D$.
By  Lemma \ref{lma:5.1}
$w=0$  in $(\R^n\setminus\Omega_0)\times(0,T)$.  
Therefore $w\big|_{\partial\Omega_0\times[0,T]}=0$
and the  restriction of the  normal derivative  of $w$ to
$\partial\Omega_0\times[0,T]$  is equal  to zero in the distribution 
sense (cf.  [E6], \S 24).  Hence 
applying the Green formula  over $D$  we get $(Lu,w)_D =
(u,L^*w)_D=0$  since  $Lu=0$  and all 
boundary terms are equal to zero.  Therefore $(u,f)=(u,L^*w)=0$ for
any  $f\in K^\perp$.  Thus $u\in K$,  i.e.
$u$  can be approximated in the norm of $V^*$  by the physically 
meaningful  solutions
$v_\e:\|u-v_\e\|_{V^*}=\max_{t\leq t\leq T}\int_{D_t}|u-v_\e|^2dx<\e$
where $\e>0$  can be chosen  arbitrary small.
\qed
 
If,  for  example,  $|u(x,t)|^2=4\sin^2\frac{\alpha}{2}+O(\e_1)$
in a small neighborhood  $U_0$  of a point $P_0$,
then  $\int_{U_0}|u(x,t)|^2dx=4\sin^2\frac{\alpha}{2}\mu(U_0)+O(\e)\mu(U_0)$,
where  $\mu(U_0)$  is the volume  of the  neighborhood 
$U_0$.
Choose  $\e$ much smaller than $\e_1$.  We get by Lemma  \ref{lma:5.2}
that
\begin{equation}                                         \label{eq:5.20}
\int_{U_0}|v_\e(x,t)|^2dx=4\sin^2\frac{\alpha}{2}\mu(U_0)+O(\e_1)\mu(U_0),
\end{equation}
i.e.
we can  determine  the flux  $\alpha$  by the measurement  of      
 a  physically meaningful solution.

\subsection{A new inverse problem for the time-dependent Schr\"odinger
equations}

Let  $(\Omega_0\times[0,T])\setminus\Omega$  be the same 
domains as before.  Let
$$
ih\frac{\partial u_p}{\partial t}-\frac{1}{2m}\sum_{j=1}^n
\Big(-ih\frac{\partial u_p}{\partial x_j}-\frac{e}{c}A_j^{(p)}(x,t)\Big)^2u_p(x,t)
-eV^{(p)}(x,t)u_p(x,t)=0
$$
be 
two Schr\"odinger equations in $(\R^n\times(0,T))\setminus \Omega$  
with  electromagneticpotentials $A^{(p)}(x,t),V^{(p)}(x,t),\ p=1,2$.  
Suppose  that supports  of $A^{(p)}(x,t)$  and
$V^{(p)}(x,t)$  are contained in $(\Omega_0\times(0,T))\setminus\Omega,\ p=1,2$.
\begin{theorem}                                     \label{theo:5.2}
Suppose

\begin{eqnarray}                                  \label{eq:5.21}
\nonumber
u_p(x,0)=f_0(x),\  x\in\R^n\setminus\Omega(0),\  p=1,2,\ 
f_0(x)=0\ \mbox{if}\ x\in\Omega_0\setminus\Omega(0),
\\
u_1(x,T)=u_2(x,T),\  x\in \R^n\setminus\Omega_0,
\end{eqnarray}
for all $f_0(x)\in L_2(\R^n\setminus\Omega_0)$.
Suppose that geometric conditions on the obstacles $\Omega(t)$ of 
 Theorems 1.1 and 1.2 in [E1] hold.
Then there exists a gauge 
$g(x,t)\in C^\infty((\R^n\times(0,T))\setminus\Omega),\ |g(x,t)|=1$
in 
$(\R^n\times(0,T))\setminus\Omega,\ g(x,t)=1$  for 
$(x,t)\in (\R^n\setminus\Omega_0)\times(0,T)$
such  that $(A^{(1)}(x,t),V^{(1)}(x,t))$ and $(A^{(2)}(x,t),V^{(2)}(x,t))$
are gauge equivalent.
\end{theorem}

{\bf Proof:}
Denote  $v(x,t)=u_1(x,t)-u_2(x,t)$.
Then  $v(x,0)=v(x,T)=0$  for  $x\in \R^n\setminus\Omega_0$  and $v(x,t)$
satisfies 
$$
ih\frac{\partial v}{\partial t}+\frac{h^2}{2m}\Delta v(x,t)=0 \ \mbox{in}\ \ 
(\R^n\setminus\Omega_0)\times(0,T).
$$
By Lemma \ref{lma:5.1} $v(x,t)=0$ in 
$(\R^n\setminus\Omega_0)\times(0,T)$.
Therefore $v\big|_{\partial\Omega_0\times(0,T)}=0$  and
$\frac{\partial v}{\partial\nu}\big|_{\partial\Omega_0\times(0,T)}=0$,
where $\frac{\partial}{\partial \nu}$ is the normal derivative.
Thus  $u_1\big|_{\partial\Omega_0\times(0,T)}= 
u_2\big|_{\partial\Omega_0\times(0,T)}$  and
$\frac{\partial u_1}{\partial\nu}\big|_{\partial\Omega_0\times(0,T)}=
\frac{\partial u_2}{\partial\nu}\big|_{\partial\Omega_0\times(0,T)}$,
By Lemma \ref{lma:5.2}  the restrictions of $u_p$  to $\partial\Omega_0\times(0,T)$ 
are dense  in $H_{-\frac{1}{2}}(\partial\Omega_0\times(0,T))$.  

Note that these restrictions exist by the partial hypoellipticity property (cf.,
for example,  [E6], \S 24). 
Therefore the Dirichlet-to-Neumann operators $\Lambda_1$ and $\Lambda_2$ are equal on   
$\partial\Omega_0\times(0,T)$.  Also $u_1(x,0)=u_2(x,0)=0$  on
$\Omega_0\setminus\Omega(0)$.  Then 
it follows from
 [E1]  that $(A^{(1)}(x,t),V^{(1)}(x,t))$ and
$(A^{(2)}(xt),V^{(2)}(x,t))$
are gauge equivalent.

{\bf Remark 5.2}  Suppose  $n=2$ and 
$$
\alpha_p=\frac{e}{h c}\int_{\partial\Omega_0}A^{(0)}(x,t)\cdot dx
$$
 are not zero, $p=1,2$.  We  assume,  in addition to (\ref{eq:5.21}),  that 
$\alpha_1=\alpha_2=\alpha$  is a constant  and the equation (\ref{eq:5.1})
has the form (\ref{eq:5.17}) for $(\R^2\setminus\Omega_0)\times(0,T)$.  
Then  Theorem \ref{theo:5.2}
holds since we can apply Lemma \ref{lma:5.1}
to the equation  (\ref{eq:5.17}) in  $(\R^2\setminus\Omega_0)\times(0,T)$.

\section{The gravitational  AB effect}
\label{section 6}
\init

\subsection{Global isometry}

We shall start with a short summary of the magnetic AB effect: consider the
Schr\"odinger equation (\ref{eq:1.1})
in $\R^2\setminus\Omega_1$  with  the boundary condition 
(\ref{eq:1.2}).  We assume that the magnetic field 
$B=\mbox{curl\ }A$  is 
zero in $\R^2\setminus\Omega_1$.  
Then locally in any simply-connected neighborhood $U\subset\R^2\setminus\Omega_1$  
the magnetic potential is gauge equivalent  to a zero potential and
$\int_\gamma A\cdot dx=0$  for a any closed curve $\gamma\subset U$.
However,  globally in $\R^2\setminus\Omega_1$ the magnetic potential
$A(x)$  may be  not gauge equivalent to a zero potential,
 in particular,  $\int_\gamma A\cdot dx=\alpha$  may  be not  zero
if $\gamma$  is a closed curve  in $\R^2\setminus\Omega_1$  encircling $\Omega_1$.

The fact that  the magnetic potential  $A(x)$  is not  gauge  equivalent globally 
in $\R_2\setminus\Omega_1$  to the zero potential  has a physical impact,  and
this phenomenon  is called  the Aharonov-Bohm effect.   More generally,
if $A_1$  and $A_2$  are not gauge  equivalent then each of them 
 makes a distinct physical impact.   Similar  situation 
(local versus global) 
appears  in  different  branches of mathematical physics.

Consider,  for example,  a pseudo-Riemannian metric
$\sum_{j,k=0}^n g_{j}(x)dx_jdx_k$  with Lorentz signature in $\Omega$,
where $x_0\in \R$  is the time variable,  $x=(x_1,...,x_n)\in\Omega,\ \Omega=
\Omega_0\setminus \cup_{j=1}^m\overline\Omega_j,\ 
\Omega_0$ is simply connected,  $\overline\Omega_j\subset\Omega_0, 
\ \Omega_j, \ 1\leq j\leq m$,
are obstacles  (cf.  subsection 3.3).  We assume that $g_{jk}(x)$  are independent
of $x_0$,  i.e.  the metric is stationary.

Consider a group  of transformations  (changes of variables)
\begin{eqnarray}                                      \label{eq:6.1}
x'=\varphi(x),
\\
x_0'=x_0+a(x),
\nonumber
\end{eqnarray}
where $x'=\varphi(x)$  is a diffeomorphism  of $\overline \Omega$  onto
$\overline{\Omega'}=\varphi(\overline \Omega)$  and
$a(x)\in C^\infty(\overline\Omega)$.  Two metrics
$\sum_{j,k=0}^n g_{jk}(x)dx_jdx_k$  and
$\sum_{j,k=0}^n g_{jk}'(x)dx_j'dx_k'$
are called isometric if
\begin{equation}                                           \label{eq:6.2}
\sum_{j,k=0}^n g_{jk}(x)dx_jdx_k=\sum_{j,k=0}^n g_{jk}'(x)dx_j'dx_k',
\end{equation}
where
$(x'_0,x')$  and  $(x_0,x)$  are related by  (\ref{eq:6.1}).

The group
of isomorphisms will play  the same role as the gauge group for
 the magnetic AB effect.

Let 
$$
\qed_gu(x_0,x)=0\ \ \mbox{in}\ \ \R\times\Omega
$$
be the wave equation corresponding to the metric $g$,   i.e.
\begin{equation}                                           \label{eq:6.3}
\qed_gu\overset{def}{=}\sum_{j,k=0}^n\frac{1}{\sqrt{(-1)^ng_0}}
\frac{\partial}{\partial x_j}
\Big(\sqrt{(-1)^n}g_0g^{jk}(x)\frac{\partial u}{\partial x_k}\Big)=0,
\end{equation}
where $g_0=\det[g_{jk}]_{j,k=0}^n,\ [g^{jk}(x)]=[g_{jk}]^{-1}$.

Solutions of (\ref{eq:6.3})  are called gravitational waves on the background of
the space-time  with the metric $g$.

Consider the initial boundary value problem for (\ref{eq:6.3}) in
$\R\times\Omega$  with  zero initial conditions
\begin{equation}                              \label{eq:6.4}
u(x_0,x)=0\ \ \mbox{for}\ x_0 \ll 0,\ x\in\Omega,
\end{equation}
and the boundary condition
\begin{equation}                               \label{eq:6.5}
u\big|_{\R\times\partial\Omega_0}=f,\ \ u\big|_{\R\times\partial\Omega_j}=0,
\ \ 1\leq j\leq m,
\end{equation}
where  $f\in C_0^\infty(\R\times\partial\Omega_0)$.  Let  $\Lambda_g$  
be the Dirichlet-to-Neumann  (DN)  operator,
i.e.  $\Lambda_g f=\frac{\partial u}{\partial \nu_g}\big|_{\R\times\partial\Omega_0}$,
where
\begin{equation}                                    \label{eq:6.6}
\frac{\partial u}{\partial \nu_g}=
\sum_{j,k=0}^ng^{jk}(x)\nu_j(x)\frac{\partial u}{\partial x_k}
\Big(\sum_{p,r=0}^ng^{pr}(x)\nu_p\nu_r\Big)^{-\frac{1}{2}}.
\end{equation}
Here 
$u(x_0,x)$  is the solution of (\ref{eq:6.3}),  (\ref{eq:6.4}),  (\ref{eq:6.5}),
$\nu(x)=(\nu_1,...,\nu_n)$  is the outward  unit normal to 
$\partial\Omega_0,\ \nu_0=0$.

Let  $\Gamma$  be an open subset  of  $\partial\Omega_0$.  We shall say that  
boundary measurements are taken  on  $(0,T)\times \Gamma$  
if we know the restriction $\Lambda_g f\big|_{(0,T)\times\Gamma}$  
for any  $f\in C_0^\infty((0,T)\times\Gamma)$.

Consider metric $g'$  in $\Omega'$ 
and the corresponding
 initial-boundary value problem
\begin{eqnarray}                                             \label{eq:6.7}
\qed_{g'}u'(x_0',x')=0\ \ \mbox{in}\ \ \R\times\Omega',
\\                                                          \label{eq:6.8}
u'(x_0',x')=0\ \ \mbox{for}\ \ x_0' \ll 0,\ x'\in \Omega',
\\                                                          \label{eq:6.9}
u\big|_{\R\times\partial\Omega_0'}=f,\ \ \ u'\big|_{\R\times\partial\Omega_j'}=0,
\ 1\leq j\leq m',
\end{eqnarray}
where 
$\Omega'=\Omega_0'\setminus\cup_{j=1}^m\overline{\Omega_j'}.$

We assume that $\partial\Omega_0\cap\partial\Omega_0'\neq \emptyset$.  Let  $\Gamma$  be
an open subset  of $\partial\Omega_0\cap\partial\Omega_0'$.

The following theorem  was proven  in [E2] (see [E2], Theorem 2.3).
\begin{theorem}                                   \label{theo:6.1}
Suppose  $g^{00}(x)>0,\ g_{00}(x)>0$  in $\overline\Omega$  and 
$(g')^{00}>0,\ g_{00}'>0$  in $\overline{\Omega'}$.  
Suppose  $\Lambda_gf\big|_{(0,T)\times\Gamma}=
\Lambda_{g'}f\big|_{(0,T)\times\Gamma}$  for  all $f\in C_0^\infty((0,T)\times\Gamma)$.
Suppose  $T>T_0$,  where $T_0$  is sufficiently large.   Then  metrics $g$  and $g'$  are isometric,  i.e.
there exists a change of variables (\ref{eq:6.1})  such that (\ref{eq:6.2})  holds.  
Moreover,  $\varphi\big|_\Gamma=I,\ a\big|_\Gamma=0$.
\end{theorem}

If two metrics $g$  and $g'$  in $\Omega$  and $\Omega'$,  respectively,  are isometric,  then the solutions $u(x_0,x)$  and $u(x_0',x')$ of 
 the corresponding wave equations are the same after 
 the change of variables (\ref{eq:6.1}).
Therefore isometric metrics have the same physical impact.

Let $g$  and $g'$ be two 
 stationary metrics in $\Omega$  and $\Omega'$,  respectively. 
Let $V$  be a neighborhood such that 
$\overline V\cap\partial\Omega_0\supset \Gamma\neq\emptyset$.
Suppose  $g$  and $g'$  are isometric in $V$,  i.e. there exists a
 change of
variables 
\begin{eqnarray}                   \nonumber
x'=\varphi_V(x),\ \ x\in \overline{V},
\\
\nonumber
x_0'=x_0+a_V(x),\ \ x\in\overline V,
\end{eqnarray}                 
such  that (\ref{eq:6.2})
holds for $x\in\overline V$.
 We want to find out what is the impact of $g$  and $g'$   being not
isometric.   
One can find a change of variables of the form (\ref{eq:6.1})  to
replace  $g'$  in $\Omega'$  by an isometric metric $\hat g$  in $\hat \Omega$ such
that $\Omega_0\cap\hat\Omega_0\supset V$  and $g=\hat g$  in $\overline V$.

It follows
from Theorem \ref{theo:6.1}
that $g$ and $\hat g$  are not isometric if and only if  the boundary measurements 
$$
\Lambda_g f\big|_{(0,T)\times\Gamma}\neq\Lambda_{\hat g}f\big|_{(0,T)\times\Gamma}\ \ \
\mbox{for some}\ \ \ f\in C_0^\infty((0,T)\times\Gamma),
$$
i.e.  metrics $g$  and $\hat g$  (and therefore $g$  and $g'$)
 have different physical impact.
This fact (i.e.   that non-isometric metric have a different  physical  impact) 
is called  the gravitational AB effect.

Note that the open set $\Gamma$ can be arbitrary small.  However the time interval 
$(0,T)$  must be large enough: $T>T_0$.

\subsection{Locally static  stationary metrics}

Let $g$ and $g'$ be isometric.  Substituting 
$dx_0'=dx_0+\sum_{j=1}^na_{x_j}(x)dx_j$  and  taking into account  that 
$dx_0$  is arbitrary,  
we get from (\ref{eq:6.1}) and (\ref{eq:6.2})  that
\begin{equation}                                \label{eq:6.10}
g_{00}'(x')=g_{00}(x),
\end{equation}
\begin{equation}                                            \label{eq:6.11}
2g_{00}'(x')\sum_{j=1}^na_{x_j}(x)dx_j+2\sum_{j=1}^ng_{j0}'(x')dx'
=2\sum_{j=1}^ng_{j0}(x)dx_j.
\end{equation}
Using (\ref{eq:6.10}) we can rewrite (\ref{eq:6.11})  in the form
\begin{equation}                                  \label{eq:6.12}
\sum_{j=1}^n\frac{1}{g_{00}'(x')}g_{j0}'(x')dx'
=\sum_{j=1}^n\frac{1}{g_{00}(x)}g_{j0}(x)dx_j-\sum_{j=1}^na_{x_j}(x)dx_j.
\end{equation}
Let $\gamma$  be an arbitrary  closed curve  in $\Omega$,  and let
$\gamma'$  be the image  of $\gamma$  in $\Omega'$   under the map
(\ref{eq:6.1}).
Integrating (\ref{eq:6.12}) we get
\begin{equation}                                  \label{eq:6.13}
\int_{\gamma'}\sum_{j=1}^n\frac{1}{g_{00}'(x')}g_{j0}'(x')dx'
=\int_\gamma\sum_{j=1}^n\frac{1}{g_{00}(x)}g_{j0}(x)dx_j,
\end{equation}
since
$\int_\gamma\sum_{j=1}^na_{x_j}(x)dx_j=0$.
Therefore the integral
\begin{equation}                                  \label{eq:6.14}
\alpha=\int_\gamma\sum_{j=1}^n\frac{1}{g_{00}(x)}g_{j0}(x)dx_j
\end{equation}
is the same for all isometric metrics.
\qed

A stationary metric $g$  is called static in $\Omega$  if  it has the form
\begin{equation}                                        \label{eq:6.15}
g_{00}(x)(dx_0)^2+\sum_{j,k=1}^n g_{jk}(x)dx_jdx_k,
\end{equation}
i.e.   when $g_{0j}(x)=g_{j0}(x)=0,\ 1\leq j\leq n,\ x\in\Omega$.

We assume that $g$ is stationary in $\Omega$  and  locally static,  i.e.
for any simply-connected neighborhood $U\subset\Omega$ there exists a change 
of variables $x'=\varphi_U(x),\ x_0'=x_0+a_U(x),\ x\in \overline U$  that
transform $g\big|_U$  to a static metric.  Therefore  (\ref{eq:6.13})
implies that
$\int_\gamma\sum_{j=1}^n\frac{1}{g_{00}(x)}g_{j0}(x)dx_j=0$
for any  closed curve $\gamma\subset U$.  
Suppose the metric $g$ is not static globally  in $\Omega$.
Then  integral (\ref{eq:6.14})  may be  not zero.  It plays
a role of  magnetic flux  for the magnetic AB effect and   $\alpha$
in (\ref{eq:6.14}) depends only  on the homotopy class of $\gamma$.

Let $V\subset\Omega,\overline V\cap\partial\Omega_0\supset\Gamma\neq\emptyset$  
and  let
$g_V'$  be a static metric  in $V$ isometric to $g$ 
 for $x\in V$.  Denote  by $g'$  an arbitrary  extension of $g_V'$  to $\overline\Omega$  such that $g'$  is static and stationary.
We have  that $g$  and $g'$  are isometric in $V$,  and we assume that
$g$  and $g'$  are not isometric in $\overline\Omega$.  
Then  the 
Theorem \ref{theo:6.1}  implies that the boundary measurements for  $g$  and  $g'$   on
$\Gamma\times(0,T)$are not  equal  for the same $f\in C_0^\infty(\Gamma\times(0,T))$.
Therefore  
$g$  and $g'$ have different physical impact,  i.e.  the gravitational AB effect  holds.
\qed

 The gravitational AB effect for a special class of locally static 
metrics was
considered previously  in [E].

\subsection{A new inverse problem for the wave equation}

 Let  $g$ and $g'$  be two stationary metrics in 
$\R^n\setminus\cup_{j=1}^m\Omega_j$  such that
\begin{equation}                                      \label{eq:6.16}
g_{jk}(x)=g_{jk}'(x) \ \ \mbox{for} \ \ |x|>R,
\end{equation}
where $R$  is large.  Assume also   that
\begin{equation}                                         \label{eq:6.17}
g_{jk}(x)=\eta_{jk}+h_{jk}(x)\ \ \mbox{for} \ \ |x|>R,
\end{equation}
where 
$$
\sum_{j,k=1}^n\eta_{jk}dx_jdx_k=dx_0^2-\sum_{j=1}^ndx_j^2
$$
is the Minkowski  metric  and $h_{jk}(x)=O(\frac{1}{|x|^{1+\e}}),\ \e>0,$
for $|x|>R$.

The following theorem is analogous to Theorem \ref{theo:5.2}.
 
\begin{theorem}                                  \label{theo:6.2}
Let $\qed_gu=0$  and $\qed_{g'}u'=0$  in $(0,T)\times(\R^n\setminus\cup_{j=1}^m\Omega_j)$,
where  $T>T_0$  (cf.  Theorem  \ref{theo:6.1}).  Consider  two  initial-boundary
value  problems
\begin{eqnarray}
\nonumber
u(0,x)=u_0(x),\ \ \ \ u'(0,x)=u_0(x),
\\
\nonumber
u_t(0,x)=u_1(x),\ \  u_t'(0,x)=u_1(x),\ \ x\in\R^n\setminus\cup_{j=1}^n\overline{\Omega_j},
\\
\nonumber
u\big|_{\partial\Omega_j\times(0,T)}=0,\ \ u'\big|_{\partial\Omega_j\times(0,T)}=0,
\ \ 1\leq j\leq m,
\\
\nonumber
u_0(x)=u_1(x)=0\ \ \mbox{in} \ B_R\setminus\cup_{j=1}^m\Omega_j,
\end{eqnarray}
where $B_R=\{x:|x|<R\}$.
Suppose  $g_{00}(x)>0,\ g_{00}'(x)>0,\ g^{00}(x)>0,\ (g')^{00}>0$  in
$\R^n\setminus\cup_{j=1}^n\Omega_j$.
If  $u_0(x)\in \overset{\circ}{H}_1(\R^n\setminus B_R)),\ u_1(x)
\in L_2(\R^n\setminus B_R)$  and if
$$
u(T,x)=u'(T,x),\ \ u_{x_0}(T,x)=u_{x_0}'(T,x),\ \ \ x\in\R^n\setminus B_R,
$$
for all $u_0(x)$  and $u_1(x)$,
then  metrics  $g$  and $g'$  are isometric  in $\R^n\setminus\cup_{j=1}^n\Omega_j$.
\end{theorem}

{\bf  Proof:}
It follows from  the existence and uniqueness theorem  that the solutions 
$u(x_0,x)$  and  $u'(x_0,x)$  belong  to 
$H_1((0,T)\times(\R^n\setminus\cup_{j=1}^n\Omega_j))$.
Let  $v=u(x_0,x)-u'(x_0,x)$.  Then 
$\qed_g v=0$  in $(0,T)\times(\R^n\setminus B_R)$
and  $v(0,x)=v_{x_0}(0,x)=0,\ v(T,x)=v_{x_0}(T,x)=0$  for  $x\in \R^n\setminus B_R$.
Extend  $v(x_0,x)$  by  zero  for $x_0>T$  and  $x_0<0$  and make the Fourier  
transform in 
$x_0: \tilde v(\xi_0,x)=\int_{-\infty}^\infty v(x_0,x)e^{-ix_0\xi_0}dx_0$.  Then 
$\tilde v(\xi_0,x)$  belongs to  $L_2(\R^n\setminus B_R)$  for  all  $\xi_0\in \R$
and  satisfies the equation
$$
L\big(i\xi_0,\frac{\partial}{\partial x}\big)\tilde v(\xi_0,x)=0,\ \ 
x\in \R^n\setminus B_R,
$$
where $L(i\xi_0,i\xi)$  is the symbol  of $\qed_g$.

It follows  from [H]  that  $\tilde v(\xi_0,x)=0$  in $\R^n\setminus B_R$  for 
all $\xi_0$.  Therefore  $u(x_0,x)=u'(x_0,x)$  for 
$x_0\in (0,T),\  x\in \R^n\setminus B_R$.
Then  $u\big|_{(0,T)\times\partial B_R}=u'\big|_{(0,T)\times\partial B_R}$
and $\frac{\partial u}{\partial\nu_g}\big|_{(0,T)\times\partial B_R}=
\frac{\partial u'}{\partial\nu_g}\big|_{(0,T)\times\partial B_R}$,
i.e.  the boundary measurements of $u$  and $u'$  on $(0,T)\times\partial B_R$
are the same. 

Analogously to the proof of Lemma \ref{lma:5.2}  one can show that
$u\big|_{(0,T)\times\partial B_R}$
and $u'\big|_{(0,T)\times\partial B_R}$  are dense 
in $H_{-\frac{1}{2}}((0,T)\times\partial B_R)$.
Hence  the DN operators $\Lambda$  and $\Lambda'$  are equal on
$(0,T)\times\partial B_R$.
 Thus Theorem \ref{theo:6.1}  implies  that  $g$  and $g'$  are isometric.
\\
\\

{\bf Acknowledgment.}

Author is grateful to Lev Vaidman of the Physics Department, Tel-Aviv University,
and Eric Hudson of the Physics Department, UCLA,
 for the stimulating discussions. 
I am thankful to Ulf Leonhardt,  St.Andrews University,   for insightful remarks.

\end{document}